\documentclass[twocolumn]{aastex61}

\usepackage{amsmath}
\usepackage{gensymb}
\usepackage{bm}
\usepackage{xcolor}
\usepackage[normalem]{ulem}   % Used for strikethrough
\usepackage{comment}

\newcommand{\sftw}[1]{\texttt{#1}}
\newcommand{\hvec}[1]{\hat{\boldsymbol{#1}}}

\definecolor{green1}{RGB}{0, 128, 0}

\newcommand{\DS}{\displaystyle}

\newcommand{\tens}[1]{\mathsf{#1}}
\newcommand{\cF}{{\cal F}}
\newcommand{\cR}{{\cal R}}
\newcommand{\cS}{{\cal S}}
\newcommand{\cU}{{\cal U}}
\newcommand{\cV}{{\cal V}}

\newcommand{\quotes}[1]{``#1''}

\DeclareMathOperator{\diag}{diag}

\shorttitle{A radiative transfer module for relativistic
magnetohydrodynamics}
\shortauthors{J. D. Melon Fuksman and A. Mignone}

\begin{document}

\title{A radiative transfer module for relativistic
magnetohydrodynamics in the PLUTO code}

\correspondingauthor{Julio David Melon Fuksman}
\email{david.melon@icranet.org}

\author{Julio David Melon Fuksman}
\affil{Dipartimento di Fisica, Sapienza Universit\`a di Roma,
 Piazzale Aldo Moro 5, 00185 Rome, Italy}
\affil{ICRANet, Piazza della Repubblica 10, 65122 Pescara, Italy}

\author{Andrea Mignone}
\affiliation{Dipartimento di Fisica, Universit\`a degli Studi di Torino, via Pietro Giuria 1, 10125 Turin, Italy}

\begin{abstract}

We present a numerical implementation for the solution of the relativistic radiation hydrodynamics and magnetohydrodynamics equations,
designed as an independent module within the freely available code \sftw{PLUTO}.
The radiation transfer equations are solved under the grey approximation and imposing the M1 closure, which allows the radiation transport to be handled in both the free-streaming and diffusion limits.
Equations are integrated following an implicit-explicit scheme,
where radiation-matter interaction terms are integrated implicitly,
whereas transport and all of the remaining source terms are solved
explicitly by means of the same Godunov-type solvers included in
\sftw{PLUTO}.
Among these, we introduce a new Harten-van Leer-contact (HLLC) solver for optically thin radiation transport.
The code is suitable for multidimensional computations in Cartesian, spherical and cylindrical coordinates, using either a single processor or parallel architectures.
Adaptive grid computations are also made possible, by means of the \sftw{CHOMBO} library.
The algorithm performance is demonstrated through a series of numerical benchmarks by investigating various different configurations with a particular emphasis on
the behavior of the solutions in the free-streaming and diffusion limits.

\end{abstract}

\keywords{radiative transfer ---  magnetohydrodynamics (MHD)
 --- relativity --- methods: numerical --- adaptive mesh refinement }

 \section{Introduction}\label{S:Introduction}
 
 Radiative transfer is of great relevance in 
 many different physical systems, 
 ocurring in a broad range of size scales.
 In the context of astrophysics, for instance, it is of fundamental
 importance in  the  modeling of star atmospheres \citep{Kippenhahn2012},
 pulsars \citep{NSandPulsars},  supernovae \citep{Fryer2007},
 and black hole accretion disks \citep{Thorne1974}.
 In some high-energy environments (e.g. gamma-ray bursts),
 matter can emit radiation while being accelerated to relativistic speeds,
 sometimes being simultaneously subject to strong electromagnetic (EM) fields
 \citep[see, e.g., the review by][]{Meszaros2006}.
 %Consequently, if such high-energy processes
 %are modeled through computational codes,
 %If all these effects are of a comparable order of importance,
 In situations where none of these effects can be disregarded,
 it is convenient to rely on numerical schemes that are able
 to deal with them simultaneously.
  
 Several numerical methods for radiation transport  found in the literature
  are based on different solutions of the
 radiative transfer equation \citep[see, e.g.,][]{Mihalas}, which provides
 a simplified yet strong formalism to the problem of radiation
 transport, absorption, and emission in presence of matter.
 This approach neglects every wave-like behavior of photons,
 and focuses, instead, on energy and momentum transport.
 Regardless of its simplicity, solving the frequency-dependent radiative
 transfer equation is not a trivial task, due to the
 high degree of nonlinearity present in the underlying mathematical
 description and the number of variables involved in it.
 For this reason, several
 simplified schemes can be found in the literature.
 Typical examples are the post-processing of ideal
 hydrodynamical calculations,
 \citep[sometimes used in cases where radiation back-reaction can be neglected,
        see, e.g.,][]{Mimica2009}, and Monte Carlo methods \citep[see e.g.][]
  {Mazzali1993, Kromer2009},
 %that, instead of attempting to solve the
 %transfer equation, compute radiation densities and fluxes
 where radiation densities and fluxes are computed
 by following the evolution of a large number
 of effective `photon packets' along selected trajectories.
 
 An alternative to these methods, followed throughout our work,
 is given by the moment approach to the radiative transfer equation.
 This consists in taking succesive angular moments of this equation, 
 in the same fashion as the hydrodynamics (HD)
 and magnetohydrodynamics (MHD) can be obtained from the collisional
 Boltzmann-Vlasov equation \citep[see e.g.][]{Goedbloed2004}.
 The resulting scheme provides an extension to relativistic 
 and non-relativistic MHD, that can be used to compute the evolution
 of the total radiation energy density and its flux, considering its
 interaction with a material fluid. 
 In this work we focus on the relativistic case, 
 to which we refer as relativistic radiation MHD (Rad-RMHD henceforth).
 The model involves a series of additional approximations
 which we now describe. 
 First, as proposed by \cite{Levermore1984}, we close the system
 of equations by assuming that 
 the radiation intensity is isotropic in a certain reference frame
 (the M1 closure).
 In addition, we consider the
 fluid to be a perfect conductor, and assume the validity of 
 the equations of ideal MHD for the interaction between matter and
 EM fields.
 Lastly, we adopt  an effective  gray body approximation,
 by replacing the opacity coefficients for a set of conveniently 
 chosen frequency-averaged values.
 
 Our implementation has been built as a supplementary module in the 
 multiphysics, multialgorithm, high-resolution code \sftw{PLUTO},
 designed for time-dependent explicit computations of either classical,
 relativistic unmagnetized or magnetized flows \citep{PLUTO}.
 The new module is fully parallel, has been adapted to all available
 geometries (Cartesian, cylindrical and spherical) and supports
 calculations on adaptively refined grids using the standard
 \sftw{PLUTO-CHOMBO} framework \citep[see][]{AMRPLUTO, CHOMBO}.
 In addition, we have introduced a new HLLC-type Riemann
 solver, suitable for optically thin radiation transport.
 In particular, our scheme is based on the HLLC solver for
 relativistic HD by \cite{MignoneBodo} and it is designed to improve the
 code's ability to resolve contact discontinuities
 when compared to HLL (Harten-van Leer) formulations
 \citep[see e.g.][]{Toro}.
 
 To integrate the transport terms of the equations of Rad-RMHD, our
 implementation employs the same sort of explicit methods used in
 \sftw{PLUTO} for the non-radiative case.
 However,
 gas-radiation interaction is treated differently, since this
 process may occur in times
 that are much shorter than the dynamical times; for instance,
 when matter is highly opaque.
 Hence, direct explicit integration of the interaction terms
 would lead to prohibitively small time steps and inefficient
 calculations.
 For this reason, our method of choice relies on Implicit-Explicit (IMEX)
 Runge-Kutta methods \citep{Pareschi2005} whereby spatial gradients
 are treated explicitly while point-local interaction terms are
 integrated via an implicit scheme.
% Such methods, commonly used to solve
% systems of differential equations with stiff source terms, 
% have a significant effect on stability
% and make it possible to integrate
% the interaction terms without having to take extremely small
% time steps.
 
 Similar approaches in the context of radiation HD and MHD
 have been followed by \citet{Gonzalez2007}, \citet{Commercon2011},
 \citet{Roedig2012}, \citet{Sadowski2013}, \citet{Skinner2013},
 \citet{Takahashi2013}, \citet{McKinney2014}
 and \citet{Rivera2016}.
 In particular, it is our intention to include our
 module in the following freely distributed versions of \sftw{PLUTO}.

 This paper is structured as follows: in Section \ref{S:RadHyd},
 we provide a brief summary of radiative transfer and the relevant
 equations used in this work, while in Section \ref{S:NumScheme}
 we give a description of the implemented algorithms. Section
 \ref{S:Tests} shows the code's performance on several selected
 tests, and in Section \ref{S:Summary} we summarize the main results
 of our work.

 \section{Radiation hydrodynamics}\label{S:RadHyd}
 
  %%%%%%%%%%%%%%%%%%%%%%%%%%%%%%%%%%%%%%%%%%%%%%%%%%%%%%%%%%%%%%%%%%%%%%
  \subsection{The equation of radiative transfer}\label{S:RadTransf}
  %%%%%%%%%%%%%%%%%%%%%%%%%%%%%%%%%%%%%%%%%%%%%%%%%%%%%%%%%%%%%%%%%%%%%%

 In this section we outline the basic derivation that leads to
 the equations of Rad-RMHD, which are described in Section \ref{S:RadRMHD}.
 We follow the formalism shown in \citet{Mihalas},
 taking as a starting point the radiative transfer equation,
 \begin{equation}\label{Eq:TransfEq}
 \begin{split}
 \frac{\partial I_\nu (t,\mathbf{x},\mathbf{n})}{\partial t} &+
 \mathbf{n} \cdot \nabla I_\nu (t,\mathbf{x},\mathbf{n})\\ &=
 \eta_\nu(t,\mathbf{x},\mathbf{n}) - 
 \chi_\nu(t,\mathbf{x},\mathbf{n})\,I_\nu (t,\mathbf{x},\mathbf{n}).
 \end{split}
 \end{equation}
 In this framework, photons are treated as point-like wave packets,
 that can be instantly emitted or absorbed by matter particles.
 As outlined in the introduction, this approach 
 rules out effects due to the wave-like nature of light such
 as diffraction, refraction, dispersion, and polarization,
 and takes care only of energy and momentum transport
 \citep[see e.g.][]{Pomraning1973}.
 Macroscopic EM fields, however, do not get such treatment
 along this work, and are instead regarded separately as
 classical fields. 
 
 Equation \eqref{Eq:TransfEq} describes the evolution of the
 radiation specific intensity $I_\nu$, defined as the amount of
 energy per unit area transported in a time interval $dt$ through
 an infinitesimal solid angle around the direction given by 
 $\mathbf{n}$, in a range of frequencies between $\nu$ and
 $\nu+d\nu$.
 The quantities on the right hand side of this equation
 describe the interaction of the gas with the radiation field. 
 The function $\eta_\nu$, known as emissivity, accounts for
 the energy released by the material per unit length,
 while the last term, proportional to $I_\nu$,
 measures the energy removed 
 from the radiation field, also per unit length.
 The total opacity $\chi_\nu$ comprises absorption and scattering
 in the medium:
 \begin{equation}
   \chi_\nu(t,\mathbf{x},\mathbf{n}) = 
   \kappa_\nu(t,\mathbf{x},\mathbf{n}) +
   \sigma_\nu(t,\mathbf{x},\mathbf{n}),
 \end{equation}
 where $\kappa_\nu$ and $\sigma_\nu$ are, respectively, the absorption
 and scattering frequency-dependent opacities.
 
 Solving Equation \eqref{Eq:TransfEq} in the presented form is
 not a trivial task since integration must be  in general
 carried out considering the dependency of $I_\nu$ on
 multiple variables $(t,\mathbf{r},\nu,\mathbf{n})$, while 
 concurrently taking into account changes in the moving material. 
 It also requires a precise
 knowledge of the functions $\eta_\nu$ and
 $\chi_\nu$, including effects such as the anisotropy
 caused by the Doppler shift.
 Instead of attempting a full solution, we adopt a
 frequency-integrated moment-based approach:
 we integrate Equation \eqref{Eq:TransfEq} over the frequency domain
 and take convenient averages in the angle -the moments- that can be
 naturally introduced in the equations of hydrodynamics.
 This procedure is described in the next section.

  %%%%%%%%%%%%%%%%%%%%%%%%%%%%%%%%%%%%%%%%%%%%%%%%%%%%%%%%%%%%%%%%%%%%%%
  \subsection{Energy-momentum conservation and interaction terms}
  \label{S:EMCons}
  %%%%%%%%%%%%%%%%%%%%%%%%%%%%%%%%%%%%%%%%%%%%%%%%%%%%%%%%%%%%%%%%%%%%%%

  We now explicitly derive the set of conservation laws describing the
  coupled evolution of fluid, EM, and radiation fields.
  While MHD quantities and radiation fields are calculated in an
  Eulerian frame of reference, absorption and scattering coefficients are
  best obtained in the fluid's comoving frame (comoving frame henceforth),
  following the formalism described in \cite{Mihalas}.
  The convenience of this choice relies on the fact that the opacity
  coefficients can be averaged easily without taking into account
  anisotropies due to a non-null fluid's 
  velocity, while the hyperbolic form of the 
  conservation equations is kept.
  In this formalism, we split the total 
  energy-momentum-stress tensor 
  $T^{\mu\nu}$ into a gas, EM, and a radiative contribution:
  \begin{equation}\label{Eq:Tmunu}
  T^{\mu\nu} = T_g^{\mu\nu} + T_{em}^{\mu\nu} + T_r^{\mu\nu}\,.
  \end{equation}
  The first of these can be written as
  \begin{equation}
  T_g^{\mu\nu} = \rho h\, u^\mu u^\nu 
  			     + p_g\,\eta^{\mu\nu},
  \end{equation}
  where $u^\mu$ is the fluid's four-velocity and $\eta^{\mu\nu}$ is
  the Minkowski tensor, while $\rho$, $h$, and $p_g$ are,
  respectively, the fluid's matter density, specific enthalpy,
  and pressure, measured in the comoving frame
  (our units are chosen so that $c=1$).
  
  This expression is valid as long as material
  particles are in \emph{local
  thermal equilibrium} (LTE henceforth), which is one of the assumptions
  of the hydrodynamical treatment.
  
  The electromagnetic contribution is given by
  the EM stress-energy tensor:
  \begin{equation}
  T_{em}^{\mu\nu} = F^{\mu\alpha} F^\nu_\alpha
  					- \frac{1}{4} \eta^{\mu\nu}F_{\alpha\beta}
  					F^{\alpha\beta}, 
  \end{equation}
  
   where the components of the field tensor $F^{\mu\nu}$
   are given by
 \begin{equation}
 F^{\mu\nu}=\begin{pmatrix}
 0 & -E_1 & -E_2 & - E_3 \\
 E_1 & 0 & -B_3 & B_2 \\
 E_2 & B_3 & 0 & -B_1 \\
 E_3 & -B_2 & B_1 & 0
 \end{pmatrix}\,,
 \end{equation} 
 where $E_i$ and $B_i$ are, respectively, the components of the
 electric and magnetic fields. 
 
  Lastly, $T_r^{\mu\nu}$ can be written in terms of the specific
  intensity $I_\nu$, as
  \begin{equation}\label{Eq:Tr}
  T_r^{\alpha\beta} = \int_0^\infty\mathrm{d}\nu 
					 \oint  \mathrm{d}\Omega\,\,
  					 I_\nu(t,\mathbf{x},\mathbf{n})\, n^\alpha n^\beta ,
  \end{equation}
  where $n^\mu \equiv (1,\mathbf{n})$ denotes
  the direction of propagation, $\mathrm{d}\nu$ the differential 
  frequency, and $\mathrm{d}\Omega$ the differential solid angle
  around $\mathbf{n}$. This expression, by definition 
  covariant \citep[see e.g.][]{Mihalas}, can be shortened as
  \begin{equation}
  T_r= \left( \begin{array}{cc}
              E_r   & F_r^i\\
              F_r^j & P_r^{ij}\\
             \end{array}    \right),
  \end{equation}
  where
  \begin{eqnarray}
  \label{Eq:RadMoments}
    E_r &\DS = \int_0^\infty\mathrm{d}\nu 
					 \oint  \mathrm{d}\Omega\,\,
  					 I_\nu(t,\mathbf{x},\mathbf{n}) \\
  					 \label{Eq:RadMomentsF}
    F_r^i&\DS = \int_0^\infty\mathrm{d}\nu 
					 \oint  \mathrm{d}\Omega\,\,
  					 I_\nu(t,\mathbf{x},\mathbf{n})\, n^i \\
    P_r^{ij} &\DS = \int_0^\infty\mathrm{d}\nu 
					 \oint  \mathrm{d}\Omega\,\,
  					 I_\nu(t,\mathbf{x},\mathbf{n})\, n^i\,n^j 	 
  \end{eqnarray}
%s
  are the first three moments of the radiation field, namely, the
  radiation energy density, the flux, and the pressure tensor. In
  our scheme, we follow separately the evolution of $E_r$ and 
  $F_r^i$,
  and define the pressure tensor in terms of these fields by means
  of a closure relation, as it is described in Section \ref{S:M1}.
  
  Following these definitions, and imposing conservation of 
  mass, total energy, and momentum, we have
  \begin{equation}
  \nabla_\mu(\rho u^\mu)=0
  \end{equation}
  and
  \begin{equation}\label{Eq:Tmunumu}
  \nabla_\mu T^{\mu\nu}=0.
  \end{equation}
  From equations \eqref{Eq:Tmunu} and \eqref{Eq:Tmunumu},
  we immediately obtain
  \begin{equation}\label{Eq:TrConsCov}
    \nabla_\mu \left( T^{\mu\nu}_{g} + T^{\mu\nu}_{em} \right)
    = -\nabla_\mu T^{\mu\nu}_{r} \equiv  G^\nu 
  \end{equation}
  where $G^\mu$ - the radiation four-force density -
  is computed by integrating Eq. \eqref{Eq:TransfEq}
  over the frequency and the solid angle, as
  \begin{equation}\label{Eq:Gmu}
  G^\mu = \int_0^\infty\mathrm{d}\nu 
					 \oint  \mathrm{d}\Omega\,\,
					 \left(
					 \chi_\nu\, I_\nu
					 -\eta_\nu
					 \right)
  					 \, n^\mu .
  \end{equation}
  The equations of Rad-RMHD can then be derived from Eq.
  \eqref{Eq:TrConsCov}, where the term $G^\mu$ accounts for
  the interaction between radiation and matter.

  The previous expression can be simplified in the comoving frame
  provided some conditions are met.
  Firstly, we assume coherent and isotropic scattering and
  calculate the total comoving emissivity as
  \begin{equation}
    \eta_\nu(t,\mathbf{x},\mathbf{n}) = 
    \kappa_\nu B_\nu(T) + \sigma_\nu J_\nu,
  \end{equation}
  where $B_\nu(T)$ is the Planck's spectral radiance
  at a temperature $T$, while $J_\nu$ is the angle-averaged
  value of $I_\nu$.
  The temperature can be determined from the ideal gas law
  \begin{equation}
  T = \frac{\mu\, m_p\, p_g}{k_B\,\rho},
  \end{equation}
  where $\mu$ is the mean molecular weight,
  $m_p$ is the proton mass, and $k_B$ the Boltzmann constant.
  
  We can then insert these expressions in Eq. \eqref{Eq:Gmu} and
  replace the opacities by their corresponding frequency-averaged values,
   such as the Planck and Rosseland means 
  \citep[see e.g.][]{Mihalas,Skinner2013}.
  In this way, we obtain the following comoving-frame source terms
  %
  %\begin{align} \label{Eq:Gc0}
  %\tilde{G}^0 &= \rho\, \kappa \left( \tilde{E}_{r} - 4\pi B(T) \right)
  % \\ \label{Eq:Gci}
  %\tilde{G}^i &= \rho\, \chi \tilde{F}_{r}^i 
  %\end{align}
  \begin{equation}\label{Eq:Gc}
    \tilde{G}^\mu = \rho \left[
               \kappa \tilde{E}_{r} - 4\pi\kappa B(T)
            ,\, \chi \tilde{\mathbf{F}}_{r} \right]
  \end{equation}
  where  $B(T)=\sigma_{\textsc{SB}}T^4/\pi c$,
  $\sigma_{\textsc{SB}}$
  is the Stefan-Boltzmann constant, and $\chi$, $\kappa$,
  and $\sigma$ are the mentioned frequency-averaged
  opacities, per unit density.
  In the code, these can either be set as constants,
  or defined by the user as functions of any set of fields
  (for instance, $\rho$ and $T$).
   From now on and except for the opacity coefficients,  we label
   with a tilde sign quantities in the comoving frame.
   
  Finally, $G^\mu$ can be obtained in the Eulerian frame
  by means of a Lorentz boost applied to Equation
  \eqref{Eq:Gc} \citep[see e.g.][]{McKinney2014}:
   \begin{equation}\label{Eq:GmuExpl}
   \begin{split}
   G^\mu= &-\kappa\rho \left(T_r^{\mu\alpha}\,u_\alpha
   + 4\pi B(T)\, u^\mu\right) \\
   & -\sigma\rho\left( T_r^{\mu\alpha}\,u_\alpha + T_r^{\alpha\beta}\,
   u_\alpha u_\beta u^\mu \right)\,.
   \end{split}
   \end{equation}
      
%   \begin{equation}\label{Eq:G0}
%   \begin{split}
%   G^0 = &-4\pi\rho\kappa\gamma B(T) 
%   		 + \rho\gamma \left(
%			\kappa - \sigma  \vert\vert \mathbf{u} \vert\vert^2 	
%   		\right) E_r \\
%   		& - \rho\left[
%			\kappa - \sigma\left(
%			\gamma^2 + \vert\vert \mathbf{u} \vert\vert^2
%			\right)   		
%   		\right]\,u_jF_r^j \\
%	    & - \rho\sigma \gamma u_j u_k P^{jk}   		
%   		\, , 
%   \end{split}
%   \end{equation}
%   \begin{equation}\label{Eq:Gi}
%   \begin{split}
%   G^i = &-4\pi\rho\kappa u^iB(T) -\rho\sigma \gamma^2 u^i E_r\\ &
%      + \rho\gamma\left[
%		\chi \delta^i_j + 2\sigma u_j u^i   
%   \right]  F_r^j \\
%   & -\rho\left(
%		\chi u_j \delta^i_k  
%		 +  \sigma u^i u_j u_k 	  \right) P^{jk}
%		 \,,
%   \end{split}   
%   \end{equation}

  %%%%%%%%%%%%%%%%%%%%%%%%%%%%%%%%%%%%%%%%%%%%%%%%%%%%%%%%%%%%%%%%%%%%%%    
  \subsection{The equations of Rad-RMHD}
  \label{S:RadRMHD}
  %%%%%%%%%%%%%%%%%%%%%%%%%%%%%%%%%%%%%%%%%%%%%%%%%%%%%%%%%%%%%%%%%%%%%%

  Assuming ideal MHD for the interaction between matter and EM fields,
  we obtain the equations of Rad-RMHD in quasi-conservative form:
  \begin{eqnarray}
    \label{Eq:RadRMHD}
    \frac{\partial \left(\mathcal{\rho\gamma}\right)}{\partial t} +
    \nabla \cdot \left(\mathcal{\rho\gamma \mathbf{v}}\right) &= 0 \\
    \frac{\partial \mathcal{E}}{\partial t} +
    \nabla \cdot \left( 
	  \mathbf{m} - \rho\gamma\mathbf{v} \right) &= G^0 \\
  \frac{\partial \mathbf{m}}{\partial t} +
  \nabla \cdot \left(
	\rho h \gamma^2 \mathbf{v}\mathbf{v}-\mathbf{B}\mathbf{B}
	-\mathbf{E}\mathbf{E} \right) + \nabla p &= \mathbf{G} \\ 
	\label{Eq:RadRMHD1a} 
  \frac{\partial \mathbf{B}}{\partial t} +
  \nabla \times \mathbf{E} &= 0 \\     \label{Eq:RadRMHD1}
  \frac{\partial E_r}{\partial t} +
  \nabla \cdot \mathbf{F}_r &= - G^0 \\
  \frac{\partial \mathbf{F}_r}{\partial t} +
  \nabla \cdot P_r &= - \mathbf{G}  ,
  \label{Eq:RadRMHD2}
  \end{eqnarray}
  where $\mathbf{v}$ is the fluid's velocity,
  $\gamma$ is the Lorentz factor, 
  $\mathbf{B}$ the mean magnetic field,
  $\mathbf{E}=-\mathbf{v}\times\mathbf{B}$ the electric field.
  In addition,  we have introduced the quantities
  \begin{equation}\label{Eq:prstot}
  p = p_g + \frac{\mathbf{E}^2+\mathbf{B}^2}{2}, 
  \end{equation}
  \begin{equation}
  \mathbf{m} = \rho h \gamma^2 \mathbf{v} + \mathbf{E}\times\mathbf{B} ,
  \end{equation}
  \begin{equation}
  \mathcal{E} = \rho h \gamma^2 - p_g - \rho\gamma +
   \frac{\mathbf{E}^2+\mathbf{B}^2}{2} ,
  \end{equation}
  which account, respectively, for the total pressure,
  momentum density, and
  energy density of matter and EM fields.
  The system \eqref{Eq:RadRMHD}-\eqref{Eq:RadRMHD2}
  is subject to the constraint
  $\nabla\cdot\mathbf{B}=0$, and
  the non-magnetic case (Rad-RHD) is recovered by taking the
  limit $\mathbf{B}\rightarrow \mathbf{0}$ in the previous expressions.
  
  In our current scheme, Equations \eqref{Eq:RadRMHD} to
  \eqref{Eq:RadRMHD2} can be solved in Cartesian, cylindrical or
  spherical coordinates.
  
  %%%%%%%%%%%%%%%%%%%%%%%%%%%%%%%%%%%%%%%%%%%%%%%%%%%%%%%%%%%%%%%%%%%%%%
  \subsection{Closure relations}\label{S:M1}
  %%%%%%%%%%%%%%%%%%%%%%%%%%%%%%%%%%%%%%%%%%%%%%%%%%%%%%%%%%%%%%%%%%%%%%

  An additional set of relations is required in order to close the
  system of Equations \eqref{Eq:RadRMHD}--\eqref{Eq:RadRMHD2}.
  An equation of state (EoS) provides closure between thermodynamical
  quantities and it can be specified as the constant-$\Gamma$ law
  \begin{equation}\label{Eq:IdealEoS}
  h = 1 + \frac{\Gamma}{\Gamma-1}\,\Theta,
  \end{equation}
  or the Taub-Mathews equation, introduced by \cite{Mathews1971},
  \begin{equation}\label{Eq:TMEoS}
  h = \frac{5}{2}\Theta + \sqrt{1+\frac{9}{4}\,\Theta^2},
  \end{equation}
  where $\Theta=p_g/\rho$.
  The properties of these equations are known and can be found, e.g., in
  \cite{MignoneMcKinney}.
  
  A further closure relation is needed for the radiation
  fields, i.e., an equation relating $P_r^{ij}$ to $E_r$
  and $\mathbf{F}_r$.
  We have chosen to implement the M1 closure, proposed by \cite{Levermore1984},
  which permits to handle both the optically thick and optically thin regimes.  
  In this scheme, it is assumed that
  $I_\nu$ is isotropic in some inertial frame, where the radiation
  stress-energy tensor takes the form
  $T'^{\mu\nu}_r=\diag(E_r',E_r'/3,E_r'/3,E_r'/3)$. 
  This leads to the following relations, which hold in any frame:
  \begin{equation}\label{Eq:M11}
  P_r^{ij}=D^{ij}E_r,
  \end{equation}
  \begin{equation}
  D^{ij}=\frac{1-\xi}{2}\,\delta^{ij}+
  \frac{3\xi-1}{2}n^in^j,
  \end{equation}
  \begin{equation}\label{Eq:M13}
  \xi=\frac{3+4f^2}{5+2\sqrt{4-3f^2}},
  \end{equation}
  where now $\bm{n}=\mathbf{F}_r/\vert\vert\mathbf{F}_r\vert\vert$
  and $f=\vert\vert\mathbf{F}_r\vert\vert/E_r$, while
  $\delta^{ij}$ is the Kronecker delta. 
  
  These relations are well behaved, as
  Equations \eqref{Eq:RadMoments} and \eqref{Eq:RadMomentsF} provide
  an upper limit to the flux, namely
  \begin{equation}\label{Eq:FsmallerE}
    \vert\vert\mathbf{F}_r\vert\vert \leq E_r,
  \end{equation}
  and therefore $0\leq f \leq 1$.

  In our scheme, we apply Equations \eqref{Eq:M11}-\eqref{Eq:M13}
  in the laboratory frame. 
  In the diffusion limit, namely,
  if $\vert\vert\mathbf{F}_r\vert\vert \ll E_r$,
  this closure leads to $P_r^{ij}=\left(\delta^{ij} \middle/ 3\right)E_r$,
  which reproduces an isotropic specific intensity known as
  Eddington limit.
  Likewise, in the free-streaming limit given by
  $\vert\vert\mathbf{F}_r\vert\vert \rightarrow  E_r$,
  the pressure tensor tends to
  $P_r^{ij}=E_r\,n^in^j$, which corresponds
  to a delta-like $I_\nu$ pointing in the same direction and
  orientation as $\mathbf{F}_r$.
  
  We point out that, even though
  both the free-streaming and the diffusion limits
  are reproduced correctly, the M1 closure may fail in some cases,
  since it implicitly assumes,  for example,
  that the intensity $I_\nu$ is axially
  symmetric in every reference frame with respect to the direction
  of $\mathbf{F}_r$.
  This is not the case, for example, when two or more
  radiation sources are involved, 
  in which case direct employment of the
  M1 closure may become inaccurate, leading to instabilities
  \citep[see e.g.][]{Sadowski2013, Skinner2013}.

 %%%%%%%%%%%%%%%%%%%%%%%%%%%%%%%%%%%%%%%%%%%%%%%%%%%%%%%%%%%%%%%%%%%%%%%
 \section{Numerical scheme}\label{S:NumScheme}
 %%%%%%%%%%%%%%%%%%%%%%%%%%%%%%%%%%%%%%%%%%%%%%%%%%%%%%%%%%%%%%%%%%%%%%%
 %\subsection{Outline of the algorithm}\label{S:ConsPrim} 
 
 For numerical purposes we write equations \eqref{Eq:RadRMHD}-\eqref{Eq:RadRMHD2}
 in conservative form as
 \begin{equation}\label{Eq:Hyp}
   \frac{\partial\cU}{\partial t} + \nabla \cdot \tens{F}(\cU)
    =  \cS(\cU)\,,
 \end{equation}
 where $\cU \equiv \left(\rho\gamma,\,\mathcal{E},\,
  \mathbf{m},\,\mathbf{B},\,E_r,\,\mathbf{F}_r  
  \right)^\intercal$ is an array of \emph{conserved} quantities,
 $\tens{F}(\cU)$ is the flux tensor and
 $\cS \equiv \left(0,G^0,\mathbf{G},
    \mathbf{0},-G^0,-\mathbf{G}\right)^\intercal$
 contains the radiation-matter interaction terms.
 The explicit expressions of $\tens{F}$ can be extracted from Equations
 \eqref{Eq:RadRMHD}-\eqref{Eq:RadRMHD2}.
 
 As outlined in the introduction, gas-radiation interaction
 may occur in timescales that are much smaller than any dynamical
 characteristic time and an explicit integration
 of the interaction terms would lead either to instabilities
 or to excessively large computing times.
 For this reason, the time discretization of Equations \eqref{Eq:Hyp}
 is achieved by means of IMEX (implicit-explicit) Runge-Kutta schemes
 \citep[see e.g.][]{Pareschi2005}.
 In the presented module, the user can
 choose between two different IMEX schemes, as
 described in Section \ref{S:IMEX}.

 In our implementation of the IMEX formalism,
 fluxes and geometrical source terms
 are integrated explicitly
 by means of standard shock-capturing Godunov-type methods, 
 following a finite volume approach.
 Fluxes are thus evaluated at cell interfaces by means of a Riemann
 solver between left and right states properly reconstructed
 from the two adjacent zones. 
 Geometrical source terms can be obtained at the cell center or following
 the approach outlined in \cite{Mig2014}.
 This \emph{explicit step} is thoroughly
 described in Section \ref{S:RSU}.
 Within this stage, we have included
 a new Riemann solver for radiation transport, which we
 introduce in Section \ref{S:HLLC}. 
 On the other hand, the integration of $G^\mu$ is performed implicitly
 through a separate step
 (the \emph{implicit step}), as described in Section \ref{S:Impl}.
 
 %%%%%%%%%%%%%%%%%%%%%%%%%%%%%%%%%%%%%%%%%%%%%%%%%%%%%%%%%%%%%%%%%%%%%%%
 \subsection{Implemented IMEX schemes}\label{S:IMEX}  
 %%%%%%%%%%%%%%%%%%%%%%%%%%%%%%%%%%%%%%%%%%%%%%%%%%%%%%%%%%%%%%%%%%%%%%%
 
 A commonly used second-order scheme is the IMEX-SSP2(2,2,2)
 method by \cite{Pareschi2005} which, when applied to
 (\ref{Eq:Hyp}), results in the following discrete scheme:
 \begin{equation} \label{Eq:IMEX1}
  \begin{array}{lcl}
   \cU^{(1)} &=& \cU^n + a\Delta t^n\cS^{(1)}
   \\ \noalign{\medskip}
   \cU^{(2)} &=& \cU^n+ \Delta t^n \cR^{(1)}  \\ 
             & &  + \Delta t^n\left[(1-2a)\cS^{(1)} + a\cS^{(2)}\right]
  \\ \noalign{\medskip}
  \cU^{n+1} &=&  \cU^n + \DS \frac{\Delta t^n}{2}\left[
 	\cR^{(1)} + \cR^{(2)}\right] \\ \noalign{\medskip}
            & & + \DS  \frac{\Delta t^n}{2}\left[ \cS^{(1)} + 
            \cS^{(2)}\right].
  \end{array}
 \end{equation}
 Here $\cU$ is an array of volume averages inside the zone $i,j,k$ (indices
 have been omitted to avoid cluttered notations), $n$ denotes the current
 step number, $\Delta t^n$ is the time step, $a=1-\sqrt{2}/2$, and the operator
 $\cR$, which
 approximates the contribution of $(-\nabla \cdot \tens{F} )$,
 is computed in an explicit fashion in terms of the conserved fields
 as detailed in Section \ref{S:RSU}.
 Potentially stiff terms
 - i.e., those poportional to $\kappa$ and $\sigma$ -
 are included in the operator $\cS$ which is solved implicitly during the
 first and second stages in Eq. (\ref{Eq:IMEX1}).

 An alternative scheme which we also consider
 in the present context is
 the following scheme (IMEX1 henceforth):
 \begin{equation} \label{Eq:IMEX2}
   \begin{array}{lcl}
   \cU^{(1)} &=& \cU^n	+ \Delta t^n \cR^{n} + \Delta t^n\,\cS^{(1)} \\
   \cU^{(2)} &=& \cU^{(1)} + \Delta t^n \cR^{(1)} +
   \Delta t^n\,\cS^{(2)} \\ 
   \cU^{n+1} &=& \DS \frac{1}{2}\left(\cU^n + \cU^{(2)}\right),
  \end{array}
 \end{equation}
 This method is an extension to the second-order
 total variation diminishing Runge-Kutta scheme (RK2) by
 \cite{GottliebShu1996}, where we have just added an
 implicit step after every flux integration.
 In the same way, we have included in the code a
 third-order version of this scheme that extends the
 third-order Runge-Kutta scheme by the same authors.
 Both the second- and third-order of this method
 are similar to those described in
 \cite{McKinney2014}.
 
 Using general methods for IMEX-RK schemes
 \citep[see e.g.][]{Pareschi2005}, it can be shown that 
 IMEX-SSP2(2,2,2) and IMEX1 are of order 2 and 1 in time and
 L- and A-stable respectively, which makes
 IMEX-SSP2(2,2,2) a seemingly better option when it comes to
 the schemes' stability. However, as we have observed
 when testing the module, the explicit addition
 of previously-calculated source terms in the last
 step of IMEX-SSP2(2,2,2) can cause inaccuracies
 whenever interaction terms are stiff and there
 are large differences in the orders of magnitude of
 matter and radiation fields
 (see Sections \ref{S:PulseThick} and \ref{S:Shadows}).
 Contrarily, IMEX1 seems
 to have better positivity-preserving properties
 and a higher accuracy in those cases.
 In general, as it is shown in Section \ref{S:Tests},
 we have obtained equivalent results with both
 methods in every other case.
 Whenever source terms can be
 neglected, both methods
 reduce to the standard RK2, which makes
 them second-order accurate
 in time for optically thin transport.

 %%%%%%%%%%%%%%%%%%%%%%%%%%%%%%%%%%%%%%%%%%%%%%%%%%%%%%%%%%%%%%%%%%%%%%%
 \subsection{Explicit step}\label{S:RSU}   
 %%%%%%%%%%%%%%%%%%%%%%%%%%%%%%%%%%%%%%%%%%%%%%%%%%%%%%%%%%%%%%%%%%%%%%%
  
  In order to compute the explicit operator $\cR$,
  %in Equation \eqref{Eq:Hyp},
  we implement a standard 
  \emph{reconstruct-solve-update} strategy
  \citep[see e.g.][]{RezzollaZanotti}.
  First, the zone averages $\cU$
  are used to compute cell-centered values of a set of
  \emph{primitive} fields, defined as
  \begin{equation}
  \cV = \left(\rho,\,p_g,\,
  \mathbf{v},\,\mathbf{B}  ,\,E_r,\,\mathbf{F}_r
  \right)^\intercal.
  \end{equation}
  Although this is a straightforward step for the radiation fields,
  as in their case primitive and conserved quantities coincide,
  this is not the case for the remaining variables.
  Primitive fields are obtained from conservative ones by means of
  a root-finding algorithm, paying
  special attention to avoiding problems related to small number
  handling that arise when large Lorentz factors are involved.
  To perform this conversion, we follow the procedure detailed
  in \cite{MignoneMcKinney}.
  
  Next, primitive fields are reconstructed to zone interfaces
  (\emph{reconstruction step}).
  In more than one dimensions, reconstruction is carried direction-wise.
  In order to avoid spurious oscillations next to
  discontinuities and steep gradients, reconstruction
  must use slope limiters in order to satisfy monotonicity constraints.
  During this step, some physical constraints are imposed,
  such as gas pressure positivity, an upper boundary for the 
  velocity given by $\vert\vert \mathbf{v} \vert\vert < 1$,
  and the upper limit to the radiation flux given by Equation
  \eqref{Eq:FsmallerE}.
    
  The reconstruction step produces left and right discontinuous states
  adjacent to zone interfaces, which we denote with $\cV_L$ and $\cV_R$.
  This poses a local initial-value problem that is solved 
  by means of an approximate Riemann solver, whose outcome
  is an estimation of the fluxes on each interface.
  In our implementation, the user can choose 
  among three of these methods.  
  The simplest one of these is
  the Lax-Friedrichs-Rusanov solver \citep[see e.g.][]{Toro}, which
  yields the following flux:
  \begin{equation}\label{Eq:LFR}
  \cF_{LF} = \frac{1}{2}\left[
  \cF_{L} + \cF_{R} -
  \vert \lambda_{max} \vert
  \left( \cU_R - \cU_L \right)
  \right].
  \end{equation}
  In this expression, $\cU_{L/R}$ and $\cF_{L/R} = \hvec{e}_d\cdot\tens{F}(\cU_{L/R})$
  are the conserved fields and flux components in the coordinate direction
  $\hvec{e}_d$  (here 
   $d=x,y,z$ in Cartesian coordinates or $d=r,\theta,\phi$ in spherical coordinates)
  evaluated at the left and right of the interface,
  while $\lambda_{\max}$ is the fastest signal speed at both sides,
  computed using both $\cV_L$
  and $\cV_R$. A less diffusive option is given by an HLL
  solver \citep[Harter-Lax-van Leer, see e.g.][]{Toro} introduced
  by \citet{Gonzalez2007}. In this case, fluxes are computed as
  \begin{equation}\label{Eq:Fhll}
     \cF_{hll} = \left\{
	       \begin{array}{ll}
	       		 \cF_L      & \mathrm{if\ } \lambda_L > 0 \\
		 \frac{\lambda_R \cF_L - \lambda_L \cF_R
		 +\lambda_R\lambda_L \left(
		 \cU_R - \cU_L
		 \right)
		 }{\lambda_R-\lambda_L} & \mathrm{if\ } 
		 	\lambda_L \leq 0 \leq \lambda_R \\
		 \cF_R     & \mathrm{if\ } \lambda_R < 0
	       \end{array}
	     \right.,
  \end{equation}
  where $\lambda_L$ and $\lambda_R$ are, respectively, the
  minimum and maximum characteristic signal speeds, taking into
  account both $\cV_L$ and $\cV_R$ states. Finally, a
  third option is given by an HLLC solver that estimates the HD (MHD) 
  fluxes as described in \cite{MignoneBodo} \citep[see also][]{MignoneBodo2006},
  and the radiation fluxes as described in Section \ref{S:HLLC}.
    
  From Eqs. \eqref{Eq:RadRMHD}-\eqref{Eq:RadRMHD2} we can
  see that, if interaction terms are disregarded, the equations
  of Rad-RMHD can be divided into two independent systems,
  one corresponding to the equations of relativistic MHD and the
  other to those of radiation transport.
  Hence, we can expect the maximum and minimum signal speeds
  of both systems to be, in the frozen limit\footnote{
  In the theory of stiff relaxation systems, the frozen limit
  refers to the small time step regime, when the effect of source
  terms on the characteristic waves is still negligible.},
  different.
  In view of this, we compute the fluxes independently for each
  subsystem of equations obtaining the speeds shown in
  Appendix \ref{S:AppSpeeds}. 
  In this manner, as it is pointed out in \cite{Sadowski2013},
  we avoid the excessive numerical diffusion that appears
  when the same signal speeds
  are used to update both radiation and MHD fields.
  This has been verified in our tests.
  
  Once the fluxes are obtained, we can compute the operator
  $\cR$ which, in the direction $d$, reads
  \begin{equation}\label{Eq:Ld}
  \cR_d(\cV)= -\frac{1}{\Delta V^d}
  \left(
	A^d_+\cF^{d}_+ -  A^d_- \cF^{d}_- 
  \right)
  +\cS_e^d,
  \end{equation}
  where $A^d_\pm$ are the cell's right ($+$) and left ($-$)
  interface areas and $\Delta V^d$ is the cell volume in that
  direction \citep[see][]{PLUTO}.
  Here $\cS^d_e(\cU)$ accounts for geometrical terms that arise when the divergence is
   written in different coordinate systems.
  The full operator $\cR$ is in the end computed as  $\sum_d \mathcal{R}_d$.

  Once the update of the conserved variables is completed, the
  time step is changed using the maximum signal speed computed
  in the previous step, according to the Courant-Friedrichs-Lewy
  condition \citep[][]{Courant1928}:
  \begin{equation}\label{Eq:Courant}
  \Delta t^{n+1}= C_a \min_d \left( 
  \frac{\Delta l_{\min}^d}{\lambda_{\max}^d} \right)
  \end{equation}
  where $\Delta l_{\min}^d$ and $\lambda_{\max}^d$ are, respectively,
  the minimum cell width and maximum signal speed along the direction
  $d$, and $C_a$, the Courant factor, is a user-defined parameter.
  %Also at this stage, the grid is updated as it is outlined
  %in Appendix \ref{S:AppC}.
 
 Finally, when magnetic fields are included, the divergence-free condition
 can be enforced using either the constrained transport method
 \citep{Balsara1999,Londrillo2004} or hyperbolic divergence cleaning
 \citep{Dedner2002,Mignone2010,Mignone2010b}.
 Both methods are available in the code.

 %%%%%%%%%%%%%%%%%%%%%%%%%%%%%%%%%%%%%%%%%%%%%%%%%%%%%%%%%%%%%%%%%%%%%%%
 \subsection{HLLC solver for radiation transport}\label{S:HLLC}   
 %%%%%%%%%%%%%%%%%%%%%%%%%%%%%%%%%%%%%%%%%%%%%%%%%%%%%%%%%%%%%%%%%%%%%%%

 We now present a novel Riemann solver for the solution
of the homogeneous radiative transfer equation.
To this purpose, we consider the subsystem formed by Eqs.
(\ref{Eq:RadRMHD1})-(\ref{Eq:RadRMHD2}) by neglecting interaction terms
and restrict our attention to a single direction,
chosen to be the $x$ axis, without loss of generality.
In Cartesian coordinates, the resulting equations take the form
\begin{equation}\label{Eq:RadHLLC}
  \frac{\partial\cU_r}{\partial t} + \frac{\partial\Phi}{\partial x}
=  0
\end{equation}
where $\cU_r = (E,\, \mathbf{F})^\intercal$ while
$\Phi = (F_x,\, P_{xx},\, P_{yx},\, P_{zx})^\intercal$
and we have omitted the subscripts $r$
for clarity purposes (we shall maintain that convention
throughout this section).
From the
analysis carried out in Appendix \ref{S:AppSpeeds}, we know that the Jacobian
$\mathbf{J}^x$ of this system has three different eigenvalues
$\{\lambda_1,\lambda_2,\lambda_3\}$, satisfying
$\lambda_1\leq\lambda_2\leq\lambda_3$.
  Since the system is hyperbolic \citep[see e.g.][]{Toro},
  the breaking of an initial discontinuity will  
  involve the development of (at most) as many waves as the
  number of different eigenvalues.
  On this basis, we have implemented
  a three-wave Riemann solver.
 
      Following \cite{HLLCradiatif}, we define the following fields:
 \begin{equation}
 \begin{array}{lcl}
 \beta_x &=&\DS \frac{3\xi-1}{2}
\frac{F_x}{\vert\vert\mathbf{F}\vert\vert^2}E \\  \noalign{\medskip}
\Pi  &=&\DS \frac{1-\xi}{2}E	\,, 
\end{array}
\end{equation}
where $\xi$ is given by Eq. \eqref{Eq:M13}.
With these definitions, the fluxes in Eq. \eqref{Eq:RadHLLC}
  can be written as
	\begin{equation}
 	\Phi=\begin{pmatrix}
           F_x \\
           F_x \,\beta_x + \Pi \\
           F_y \,\beta_x \\
           F_z \,\beta_x
         \end{pmatrix},
 	\end{equation}
 	and $F_x$ can be shown to satisfy $F_x=\left(E+\Pi\right)\beta_x$.
 	These expressions are similar to those of relativistic hydrodynamics
 	(RHD henceforth),
 	where $\beta_x$, $\Pi$ and $\mathbf{F}$ play, respectively, the role of $v_x$, $p_g$
 	and $\mathbf{m}$ while $E$ is tantamount to the total energy.
        With the difference that there is no field corresponding to density,
        the equations are exactly the
 	same as those corresponding to energy-momentum conservation of a fluid,
 	with a different closure relation.
 	
 	With this in mind, we follow
 	analogous steps to those in \cite{MignoneBodo}
 	in order to construct a HLLC solver
 	for the system defined by Equations \eqref{Eq:RadHLLC}.
 	In this case,
 	instead of the intermediate constant state considered in the HLL solver,
 	we include an additional
        middle wave (the analog of a \quotes{contact} mode)
        of speed $\lambda^*$ that separates
 	two intermediate states $\mathcal{U}^*_L$ and $\mathcal{U}^*_R$, where
 	\begin{equation}\label{Eq:CondLambdaS}
 	\lambda_L\leq\lambda^*\leq\lambda_R\,.
 	\end{equation}
 	In this way, the full approximate
 	solution verifies
    \begin{equation}
      \cU_r(0,t)=\begin{cases}
      \cU_{r,L}   &  \text{if } \lambda_L> 0 \\
      \cU^*_{r,L} &  \text{if } \lambda_L\leq 0 \leq \lambda^* \\
      \cU^*_{r,R} &  \text{if } \lambda^*\leq 0 \leq \lambda_R \\
      \cU_{r,R}   &  \text{if } \lambda_R< 0 \,.
	\end{cases}
    \end{equation}     
    The corresponding fluxes are
    \begin{equation}
    \Phi_{hllc}(0,t)=\begin{cases}
    \Phi_L   &  \text{if } \lambda_L> 0 \\
    \Phi^*_L &  \text{if } \lambda_L\leq 0 \leq \lambda^* \\
    \Phi^*_R &  \text{if } \lambda^*\leq 0 \leq \lambda_R \\
    \Phi_R   &  \text{if } \lambda_R< 0 \,.
	\end{cases}
    \end{equation}
    States and fluxes are related by the Rankine-Hugoniot jump conditions across
    the outermost waves $\lambda_S$ ($S=L,R$),
    \begin{equation}\label{Eq:RH_Rad}
      \lambda_S\,(\cU^*_{r,S} - \cU_{r,S}) = \Phi^*_S - \Phi_S\,.
    \end{equation}
    A similar condition must also hold across the middle wave so that,
    when Equation \eqref{Eq:RH_Rad} is applied to all three waves, one has at
    disposal a system of 12 equations for the 17 unknowns
   ($\cU^*_{r,L}$, $\cU^*_{r,R}$, $\Phi^*_L$,
    $\Phi^*_R$, and $\lambda^*$) and therefore further assumptions
    must be made.
    From the results of the tests performed with the HLL 
    solver, we have verified  that $\beta_x$ and $\Pi$
    are conserved along the intermediate contact
    mode for all the obtained solutions.
    Noting that $\lambda_2(E,\mathbf{F})=\beta_x(E,\mathbf{F})$,
    it can be seen that, for a discontinuity of speed $\beta_x$
    along which $\beta_x$ and $\Pi$ are continuous, the
    jump conditions
    \eqref{Eq:RH_Rad} are satisfied, as pointed out in
    \cite{HLLCradiatif} and proven in \cite{Hanawa2014}.
	Thus, we impose the constraints
    $\lambda^*=\beta^*_{x,L}=\beta^*_{x,R}$ and
    $\Pi^*_L=\Pi^*_R$.
    These conditions are analogous to those satisfied by the 
    contact discontinuity in RHD, across which $p_g$ and $v_x$
    are conserved, and where the latter
    coincides with the propagation speed.
    Following \cite{MignoneBodo}, we assume that
    $\Phi^*$ can be written in terms of
    the five variables $(E^*,\Pi^*,\beta_x^*,F^*_y,F^*_z)$
    in the following way:
	\begin{equation}
 	\Phi^*=\begin{pmatrix}
           F^*_x \\
           F^*_x \,\beta^*_x + \Pi^* \\
           F^*_y \,\beta^*_x \\
           F^*_z \,\beta^*_x
         \end{pmatrix},
 	\end{equation}
 	where for consistency we have defined
 	$F^*_x\equiv(E^*+\Pi^*)\beta^*_x$.
 	Under these constraints, the jump conditions across the middle
	wave are automatically satisfied, and Eq. \eqref{Eq:RH_Rad}
 	is reduced to the following system of 8 equations
 	in 8 unknowns:
	\begin{equation}\label{Eq:RH_Rad1}
	\begin{array}{lcl}
	E^*(\lambda-\lambda^*)&=&
	E(\lambda-\beta_x)+\Pi^*\lambda^*-\Pi\,\beta_x\\ 
	F_x^*(\lambda-\lambda^*)&=&F_x(\lambda-\beta_x)+\Pi^*-\Pi \\
	F_y^*(\lambda-\lambda^*)&=&F_y(\lambda-\beta_x) \\ 
	F_z^*(\lambda-\lambda^*)&=&F_z(\lambda-\beta_x)\,,
	\end{array}
	\end{equation}	 	
    which holds for both subscripts L and R (we shall maintain
    this convention in what follows). The first two
    equations in Eq. \eqref{Eq:RH_Rad1}
    can be turned into the following quadratic
    expression, from which $\lambda^*$ can be obtained:
	\begin{equation}\label{Eq:PLPR}
	(A_L\lambda^*-B_L)(1-\lambda_R\lambda^*)=
	(A_R\lambda^*-B_R)(1-\lambda_L\lambda^*),
	\end{equation}
	with
	\begin{align}\label{Eq:Adef}
	A &= \lambda E - F_x \\\label{Eq:Bdef}
	B &= (\lambda - \beta_x) F_x - \Pi.
	\end{align}
	Once $\lambda^*$ is known, we can compute $\Pi^*$ as
	\begin{equation}
	\Pi^*=\frac{A\,\lambda^*-B}{1-\lambda\,\lambda^*},
	\end{equation}
	and the remaining fields from Eq. \eqref{Eq:RH_Rad1}.
    Similarly to the RHD counterpart, among the
	two roots of Equation \eqref{Eq:PLPR} we must choose the only
	one that guarantees $\lambda^*\in[-1,1]$, which in our case
	corresponds to that with the minus sign.
    As shown in Appendix \ref{S:AppLambdaS}, this definition
    of $\lambda^*$ satisfies Eq. \eqref{Eq:CondLambdaS}.
    We have also checked by means of extensive numerical testing
    that the intermediate states $\mathcal{U}^*_L$ and $\mathcal{U}^*_R$
    constructed in this way satisfy Equation \eqref{Eq:FsmallerE},
    which guarantees the positivity of our HLLC scheme.
    However, unlike the RHD case, the coefficients
    $\{A_L,B_L,A_R,B_R\}$
    defined in Equations \eqref{Eq:Adef} and \eqref{Eq:Bdef}
    can simultaneously be equal to zero, meaning that
    $\lambda^*$ can no longer be determined from Equation \eqref{Eq:PLPR}.
    This happens under
    the conditions $\vert\vert\mathbf{F}\vert\vert = E$ for both L and R,
    and $F_{xL}/\vert\vert\mathbf{F}_L\vert\vert\leq
    F_{xR}/\vert\vert\mathbf{F}_R\vert\vert$, in which case the jump
    conditions lead to the formation of vacuum-like intermediate
    states.
    We overcome this issue by switching the solver to the
    standard HLL whenever these conditions are met.
    
    As for the HLL solver, signal velocities must be limited when
    describing radiation transfer in highly opaque materials in order
    to reduce numerical diffusion (see Appendix \ref{S:AppSpeeds}).
    Whenever this occurs, we also switch
    to the standard HLL solver,
    and limit $\lambda_L$ and $\lambda_R$ according to
    Equation \eqref{Eq:RadSpeedLim}. Hence, we can only expect the 
    HLLC solver to improve the accuracy of the obtained solutions in
    optically thin regions of space, whereas the results should be
    the same for both HLL and HLLC everywhere else. Finally, although
    the use of the HLLC solver can reduce the numerical diffusion
    when compared to the HLL solver, this can cause
    spurious oscillations around shocks
    that would be damped with a more diffusive method. As for the HLLC
    solver for relativistic HD and MHD included in \sftw{PLUTO}, this
    problem can be reduced by implementing an additional flattening
    in the vicinity of strong shocks \citep[see e.g.][]{MignoneBodo}.
     
 %%%%%%%%%%%%%%%%%%%%%%%%%%%%%%%%%%%%%%%%%%%%%%%%%%%%%%%%%%%%%%%%%%%%%%%
   \subsection{Implicit step}\label{S:Impl}
 %%%%%%%%%%%%%%%%%%%%%%%%%%%%%%%%%%%%%%%%%%%%%%%%%%%%%%%%%%%%%%%%%%%%%%%

   We now describe the algorithm employed for the
   implicit integration of the radiation-matter interaction
   terms.
   A typical implicit step of an IMEX scheme 
   (see Eqs. \ref{Eq:IMEX1} and \ref{Eq:IMEX2})
   takes the form
   \begin{equation} \label{Eq:ImplEq0}
   \cU = \cU' + s\, \Delta t^n\,\cS\,,
   \end{equation}
   where $s$ is a constant and primed terms denote
   some intermediate state value.
   Equation \eqref{Eq:ImplEq0} shows that
   the mass density, computed as $\rho\gamma$, as well as
   the total energy and momentum densities, defined as
   $ E_{tot} = \mathcal{E} + E_r$ and
   $\mathbf{m}_{tot}=\mathbf{m} + \mathbf{F}_r$, must be
   conserved during this partial update owing
   to the particular form of the source terms.
   This yields the following implicit relations between $\mathcal{V}$
   and $\cU_r$:
   \begin{align}\label{Eq:PrimImpl1}
   \begin{array}{ll}
   \mathcal{E}(\mathcal{V}) &= E_{tot}-E_r \\
   \mathbf{m}(\mathcal{V}) &= \mathbf{m}_{tot}-\mathbf{F}_r .   
   \end{array}
   \end{align}
   We can then solve Eq. \eqref{Eq:ImplEq0} in terms of the
   following reduced system:
   \begin{equation} \label{Eq:ImplEq}
   \cU_{r} = \cU'_{r} - s\, \Delta t^n\,\mathcal{G}\,,
   \end{equation}    
   with
   % $\mathcal{U}_r \equiv (E_r,\mathbf{F}_r)^\intercal$ and
   $\mathcal{G} \equiv (G^0,\mathbf{G})^\intercal$,
   where $G^\mu$ is given in Eq. \eqref{Eq:GmuExpl}.
   In Eq. \eqref{Eq:ImplEq}, radiation fields can be regarded
   as functions of the MHD fields and vice-versa by means of Eq.
   \eqref{Eq:PrimImpl1}, and therefore the system can be solved
   in terms of either one of these.
   
   In order to solve Equation \eqref{Eq:ImplEq}, we have implemented
   and compared three different multidimensional root finder
   algorithms, which we now describe.

   \begin{enumerate}
   \item \emph{Fixed-point method}.
   This method \citep[originally proposed by][]{Takahashi2013} is based
   on iterations of $\cU_r$ and follows essentially the same
   approach outlined by \cite{Palenzuela2009} 
   in the context of resistive MHD.
   In this scheme all of the MHD primitive 
    variables, as well as $D^{ij}$, are written at a previous
    iteration with respect to $\cU_r$.
    In that manner, 
    $\mathcal{G}$ can be written at a given iteration $m$ as
    \begin{equation}
    \mathcal{G}^{(m)}=
    \mathcal{M}^{(m)}\cU^{(m+1)}_r+b^{(m)} ,
    \end{equation}
    where
    $\mathcal{M}$ is a matrix and $b$ a column vector,
    both depending on $\mathcal{V}$ and $D^{ij}$, and
    the numbers between parentheses indicate the iteration in which
    the fields are evaluated. Inserting this in Equation
    \eqref{Eq:ImplEq}, the updated conserved fields can be computed
    as
    \begin{equation}   
    \cU^{(m+1)}_r = 
    \left( \mathcal{I} + s\, \Delta t^n \mathcal{M}^{(m)} \, \right)^{-1}
    \left( \cU'_r - s\, \Delta t^n\, b^{(m)} \right),
    \end{equation}
    after which primitive fields can be updated using Eq.
    \eqref{Eq:PrimImpl1}.
    
    \item \emph{Newton's method for radiation fields},
    implemented in \cite{Sadowski2013} and \cite{McKinney2014}.
    This scheme consists in finding the roots of the nonlinear
    multidimensional function
    \begin{equation}
    \mathcal{Q}(E_r,\mathbf{F}_r)
    =\cU_r-\cU'_r+ s\,\Delta t^n\, \mathcal{G},
    \end{equation}

    updating the radiation variables on each iteration as
    \begin{equation}
    \cU_r^{(m+1)}=\cU_r^{(m)}
    -\left[\mathcal{J}^{(m)}\right]^{-1}
    \mathcal{Q}^{(m)},
    \end{equation}
    where we have defined the Jacobian matrix $\mathcal{J}$ as
    $\mathcal{J}_{ij}=\partial \mathcal{Q}_i / \partial \cU_r^j$. 
    The elements of $\mathcal{J}$ are computed numerically, taking
    small variations of the iterated fields.
    As in the fixed-point method,
    matter fields are computed from $\cU_r$ for each step
    by means of an inversion of
    Eq. \eqref{Eq:PrimImpl1}.
    
    \item \emph{Newton's method for matter fields},
    implemented in \cite{McKinney2014}.
    This procedure is identical to the previous one,
    with the difference that in this case the iterated fields are
    the fluid's
    pressure and the spatial components of its four-velocity,
    which we denote as $\mathcal{W}=(p_g,\mathbf{u})^\intercal$.
    These are updated as
    \begin{equation}
    \mathcal{W}^{(m+1)}=\mathcal{W}^{(m)}
    					-\left[\mathcal{J}^{(m)}\right]^{-1}
    					\mathcal{Q}^{(m)},
    \end{equation}
    where now $\mathcal{J}_{ij}=\partial \mathcal{Q}_i /
    \partial \mathcal{W}^j$
    and $\mathcal{Q}$ is regarded as a function of $\mathcal{W}$. 
    This scheme is much faster than the previous one, since
    the computation of $\cU_r$ from $\mathcal{W}$
    by means of Eq. \eqref{Eq:PrimImpl1} 
    is now straightforward, and no longer requires
    a cumbersome inversion of conserved to primitive fields.
    
    \end{enumerate}
   
       For each of these methods, 
	iterations are carried out until convergence is reached
       by means of some error function.
       In the first
	of them, such function is chosen as the norm of the relative
	differences between successive values of $\mathcal{V}$, whereas
	in the last two of them it is defined as the norm of
	$\mathcal{Q}^{(m+1)}$.
	If $\mathcal{E}\ll E_r$,
	the	errors of the matter fields can be large even when
	radiation fields converge, since Eq.
	\eqref{Eq:PrimImpl1}
	implies that $\mathcal{E}$ and $E_r$ have the same absolute
	error, as well as $\mathbf{m}$ and $\mathbf{F}_r$.
    Therefore,
	having small relative differences of $E_r$ does not guarantee
	the same for $\mathcal{E}$, which can lead to non-negligible
	inaccuracies if the second method is used.
	Equivalently, the same problem can occur whenever
	$\mathcal{E}\gg E_r$
	if method 3 is chosen \citep[see also][]{McKinney2014}.
	To overcome this issue, we have included in the code the
	option of adding to the convergence function the norm of the
	relative differences of $\mathcal{E}$ when using the
	second method, and of $E_r$ when using the third one.
	We have seen in the performed tests that the fixed-point
	method converges rather fast, meaning that
	the number of iterations
	that it requires frequently coincides with that obtained
	with the last two methods. This scheme has sufficed to perform
	all the tests carried out in this work, being often the fastest
	one when compared to the other two, having been overcome only
	by method 3 in a few cases.

%%%%%%%%%%%%%%%%%%%%%%%%%%%%%%%%%%%%%%%%%%%%%%%%%%%%%%%%%%%%%%%%%%%%%%%%
 \section{Numerical Benchmarks}\label{S:Tests}
%%%%%%%%%%%%%%%%%%%%%%%%%%%%%%%%%%%%%%%%%%%%%%%%%%%%%%%%%%%%%%%%%%%%%%%%
 
 We show in this section a series of  numerical benchmarks to verify
 the code performance, as well as the correctness of the implementation
 under different physical regimes and choices of coordinates.
 Unless otherwise stated, we employ the HLLC 
 solver introduced in Section \ref{S:HLLC}, the
 domain is discretized using a fixed uniform grid and
 outflow boundary conditions are imposed for all the fields. 
 Magnetic fields are neglected in all the considered problems,
 except in Section \ref{S:TestRMHD}.
 Furthermore, all the
 tests have been run with both the IMEX-SSP2(2,2,2) and IMEX1 methods,
 obtaining equivalent results unless indicated otherwise.
 
%%%%%%%%%%%%%%%%%%%%%%%%%%%%%%%%%%%%%%%%%%%%%%%%%%%%%%%%%%%%%%%%%%%%%%%%
 \subsection{Riemann Problem for optically-thin radiation transport}
 \label{S:ShockThin}
%%%%%%%%%%%%%%%%%%%%%%%%%%%%%%%%%%%%%%%%%%%%%%%%%%%%%%%%%%%%%%%%%%%%%%%%

 We first validate the implemented radiation transport schemes when
 any interaction with matter is neglected.
 To this end, we have run several
 one-dimensional Riemann problems setting all the interaction terms to
 zero and focusing only on the evolution of the radiation
 fields. The initial setup of these consists of
 two regions of uniform $E_r$ and $\mathbf{F}_r$, separated by a
 discontinuity at $x=0$.
 The full domain is defined as the interval $[-20,20]$.
% In these tests,
% we have  also  considered perpendicular fluxes along the $y$ %direction.
 We show here two of such tests, exploring the case 
 $\vert\vert \mathbf{F}_r \vert\vert < E_r$ (test 1) and the free-streaming 
 limit, $\vert\vert \mathbf{F}_r \vert\vert \simeq E_r$ (test 2).

% \begin{figure}[t!]
%	\centering
%	\includegraphics[scale=0.65]{plots/Paper_OptThinShocks_test4}
%	\caption{ Optically thin Riemann test 1.
%	From top to bottom, $E_r$,
%	the $x$ and $y$ components of $\mathbf{F}_r$, and the fields
%	$\Pi$ and $\beta_x$ defined in Section \ref{S:HLLC} are plotted
%	at $t = 20$. This solution, computed with a resolution of
%	$2^{14}$ zones, shows a left shock at $x\approx-11$, a right
%	rarefaction wave at $x\approx 11$, and a central contact
%	discontinuity at $x\approx 1$. It can be seen that $\Pi$ and
%	$\beta_x$ are continuous along the last of these. }
%	\label{fig:OptThinShocks1}
%  \end{figure}

   \begin{figure}[t!]
	\centering
	\includegraphics[width=0.47\textwidth]{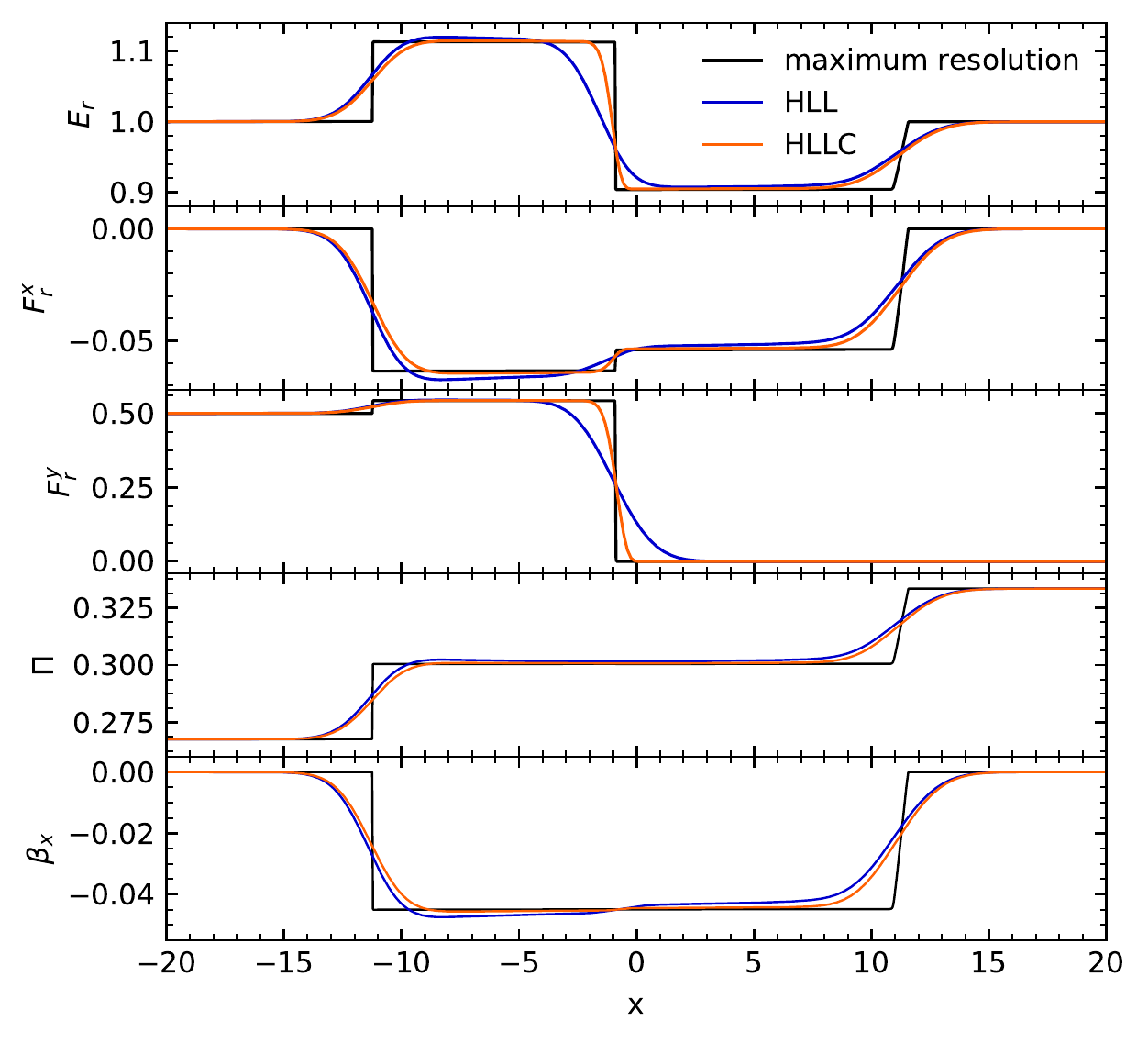}
	\caption{ Radiation fields in the optically thin Riemann
	test 1 at $t=20$. Two solutions obtained with
	the HLL solver (solid blue line) and the HLLC solver
	(solid orange line), computed
	using $2^8$ zones in both cases,
	 are compared to a reference solution
	 obtained with $2^{14}$ zones.
	These show a left shock at $x\approx-11$, a right
 	expansion wave at $x\approx 11$, and a central contact
 	discontinuity at $x\approx -1$, along which the fields $\Pi$ and
 	$\beta_x$ are continuous.}
	\label{fig:Paper_hll_hllc_test1}
  \end{figure}  
  
     \begin{figure}[t!]
	\centering
	\includegraphics[width=0.47\textwidth]{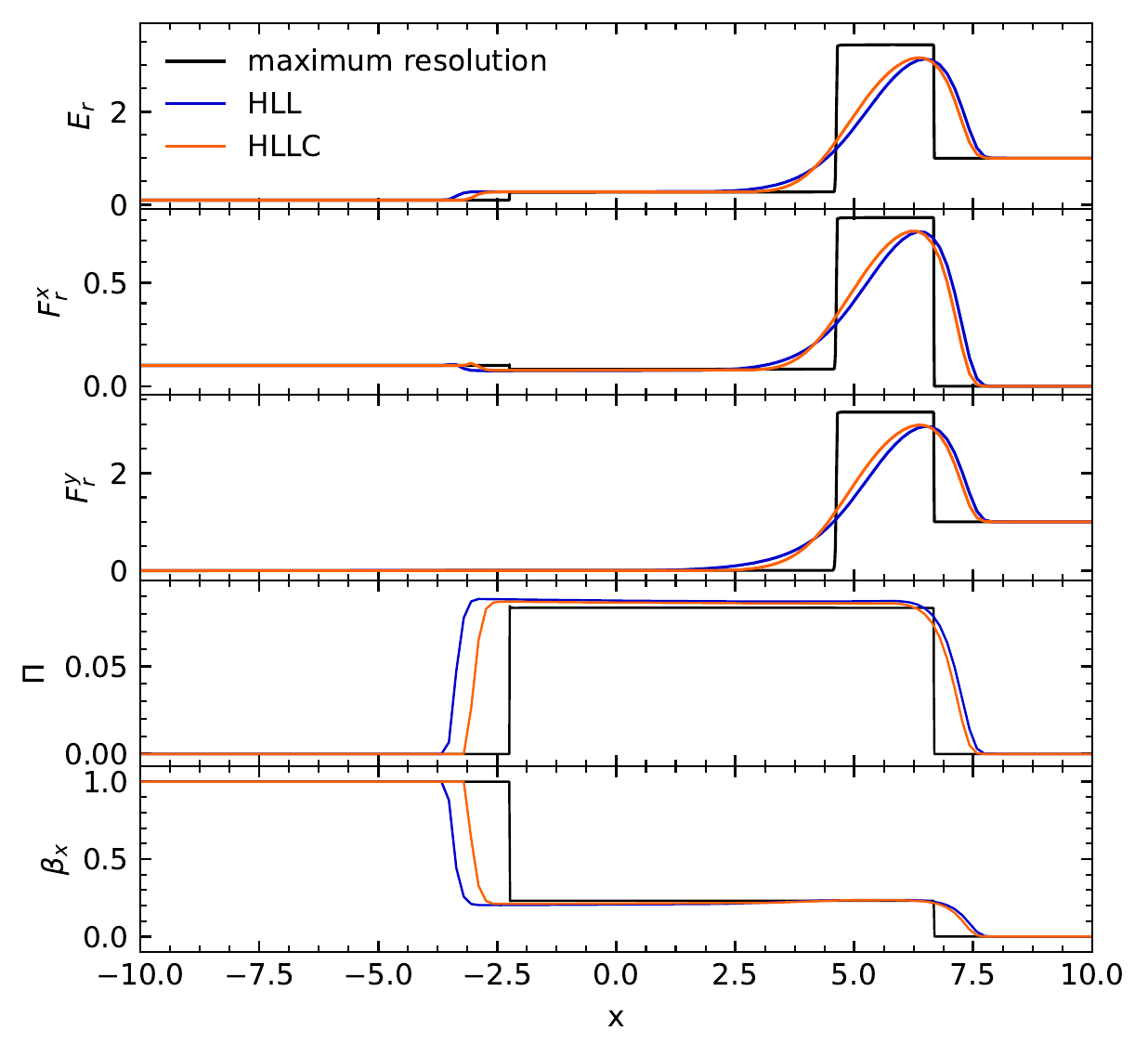}
	\caption{ Same as Fig. \ref{fig:Paper_hll_hllc_test1},
	for the optically thin Riemann test 2.
	The solutions exhibit a leftward-moving shock,
	a contact discontinuity and a rightward-moving shock,
	at $x\approx -2.2$, $4.5$ and $7$ respectively. }
	\label{fig:Paper_hll_hllc_test2}
  \end{figure}  
 
 In the first test, initial states are assigned at $t=0$ as
\begin{equation}
  (E_r,\, F_r^x,\, F_r^y)_{L,R} = \left\{\begin{array}{ll}
  \left(1,\, 0,\, \frac{1}{2}\right)   & \;{\rm for}\quad x < 0  \\ \noalign{\medskip}
  \left(1,\, 0,\, 0\right)     & \;{\rm for}\quad x > 0  
  \end{array}\right.
\end{equation}
 The solution, plotted in Fig \ref{fig:Paper_hll_hllc_test1} at $t=20$
 with a resolution of $2^{14}$ zones (solid black line),
 shows a three-wave pattern as it is expected from the eigenstructure
 of the radiation transport
 equations (see Section \ref{S:HLLC} and Appendix \ref{S:AppLambdaS}).
 The left and right outermost waves are, respectively, a left-facing shock
 and a right-going expansion wave, while the middle wave
 is the analog of a contact wave. 
 The fields $\Pi$ and $\beta_x$, defined in Section \ref{S:HLLC},
 are constant across the contact mode.
 On the same Figure, we show the solution obtained with the HLL and HLLC solvers
 at the resolution of 256 zones using a $1^{\rm st}$ order
 reconstruction scheme \citep[see][]{MignoneBodo}.
 As expected, the employment of the HLLC solver yields
 a sharper resolution of the middle wave. 

 For the second test, the initial condition is defined as
\begin{equation}
  (E_r,\, F_r^x,\, F_r^y)_{L,R} = \left\{\begin{array}{ll}
   \left(\frac{1}{10},\, \frac{1}{10},\, 0\right) & \;{\rm for}\quad x < 0  \\ \noalign{\medskip}
   \left(     1     ,\,  0,\, 1 \right)    & \;{\rm for}\quad x > 0  
  \end{array}\right.
\end{equation}
 Results obtained with the $1^{\rm st}$-order scheme and the HLL and HLLC solvers
 are plotted in Fig. \ref{fig:Paper_hll_hllc_test2} together with the reference
 solution (solid black line) at $t=20$.
 As for the previous case, a three-wave pattern emerges,
 formed by two left- and right-going shocks
 and a middle contact wave.
 It can be also seen that $\Pi$ and $\beta_x$
 are again continuous across the contact wave.
 Differences between HLLC and HLL are less pronounced than the previous case,
 with the HLL (HLLC) overestimating the left-going shock position
 by 50\% (30\%).

\begin{figure}[t!]
	\centering
	\includegraphics[width=0.47\textwidth]{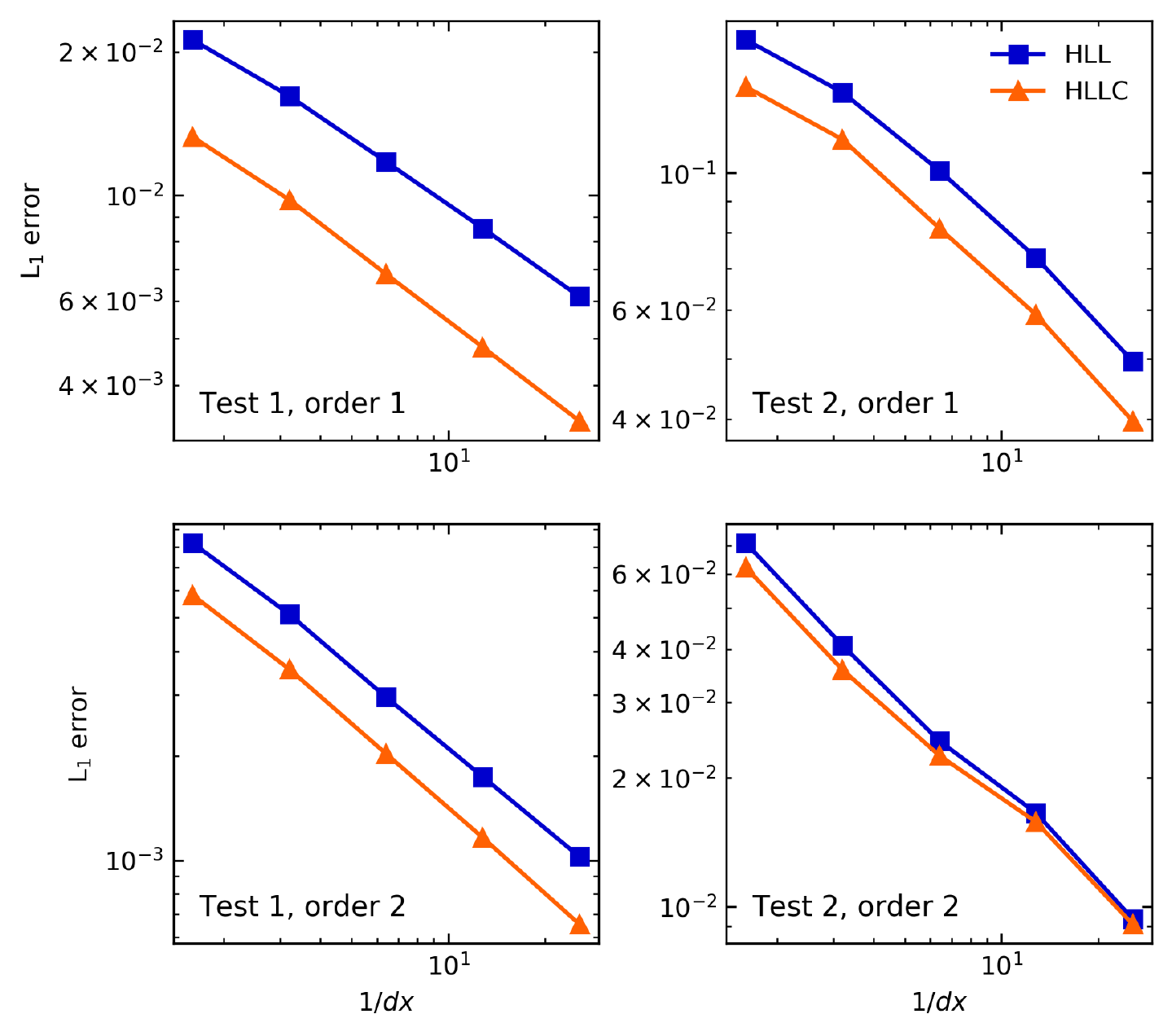}
	\caption{ L$_1$ error of $E_r$ in
	the optically thin Riemann tests
	1 and 2, computed in each case
	with respect to a reference solution
	obtained using $2^{14}$ zones. The errors are plotted
	for several resolutions as a function of $1/dx$, where
	$dx$ is the cell's width in each case. Different results
	are shown using first-order (upper panels) and
	second-order (lower panels) reconstruction schemes.
}
	\label{fig:Paper_err_hll_hllc}
  \end{figure}

 For both tests, we have conducted a resolution study covering the
 range $[2^{6},2^{10}]$ using $1^{\rm st}$- as well as $2^{\rm nd}$-order
 reconstructions making use of the second-order harmonic mean limiter by
 \cite{vanLeer1974}.
 In Figure \ref{fig:Paper_err_hll_hllc}, we plot the error in
 L$_1$-norm of $E_r$ (computed with respect to the reference solution)
 as functions of the resolution.
 The Courant number is $C_a = 0.4$ for both cases.
 Overall, the HLLC yields smaller errors when compared to HLL,
 as expected.
 This discrepancy is more evident in the $1^{\rm st}-$order case and it
 is mitigated in the case of a $2^{\rm nd}$ order interpolant
 \citep[a similar behavior is also found in][]{MignoneBodo}.

%%%%%%%%%%%%%%%%%%%%%%%%%%%%%%%%%%%%%%%%%%%%%%%%%%%%%%%%%%%%%%%%%%%%%%%%
 \subsection{Free-streaming beam}\label{S:Beams}
%%%%%%%%%%%%%%%%%%%%%%%%%%%%%%%%%%%%%%%%%%%%%%%%%%%%%%%%%%%%%%%%%%%%%%%%
   
 A useful test to investigate the code's accuracy for multidimensional
 transport is the propagation of a radiation beam oblique 
 to the grid \citep[see e.g.][]{Richling2001,Gonzalez2007}.
 This problem is also useful to quantify the numerical diffusion
 that may appear when fluxes are not aligned with the axes.
 We again neglect the gas-radiation interaction terms,
 and follow solely the evolution of the radiation fields.

 The initial setup consists of a square Cartesian
 grid of side $L=5$ cm, where the radiation energy density is set to 
 $E_{r,0}=10^{4}$ \mbox{erg cm$^{-3}$}.
 At the $x=0$ boundary, a radiation beam is injected by fixing 
 $E_r=10^8\,E_{r,0}$ and $\mathbf{F}_r=(1/\sqrt{2},1/\sqrt{2})\, E_r$
 for $y\in [0.30,0.44]$ cm.
 Thus, the injected beam satisfies the equality
 $\vert\vert\mathbf{F}_r\vert\vert=E_r$, which
 corresponds to the free-streaming limit.
 Outflow conditions are imposed on the remaining boundaries.

\begin{figure}[t]
	\centering
	\includegraphics[width=0.48\textwidth]{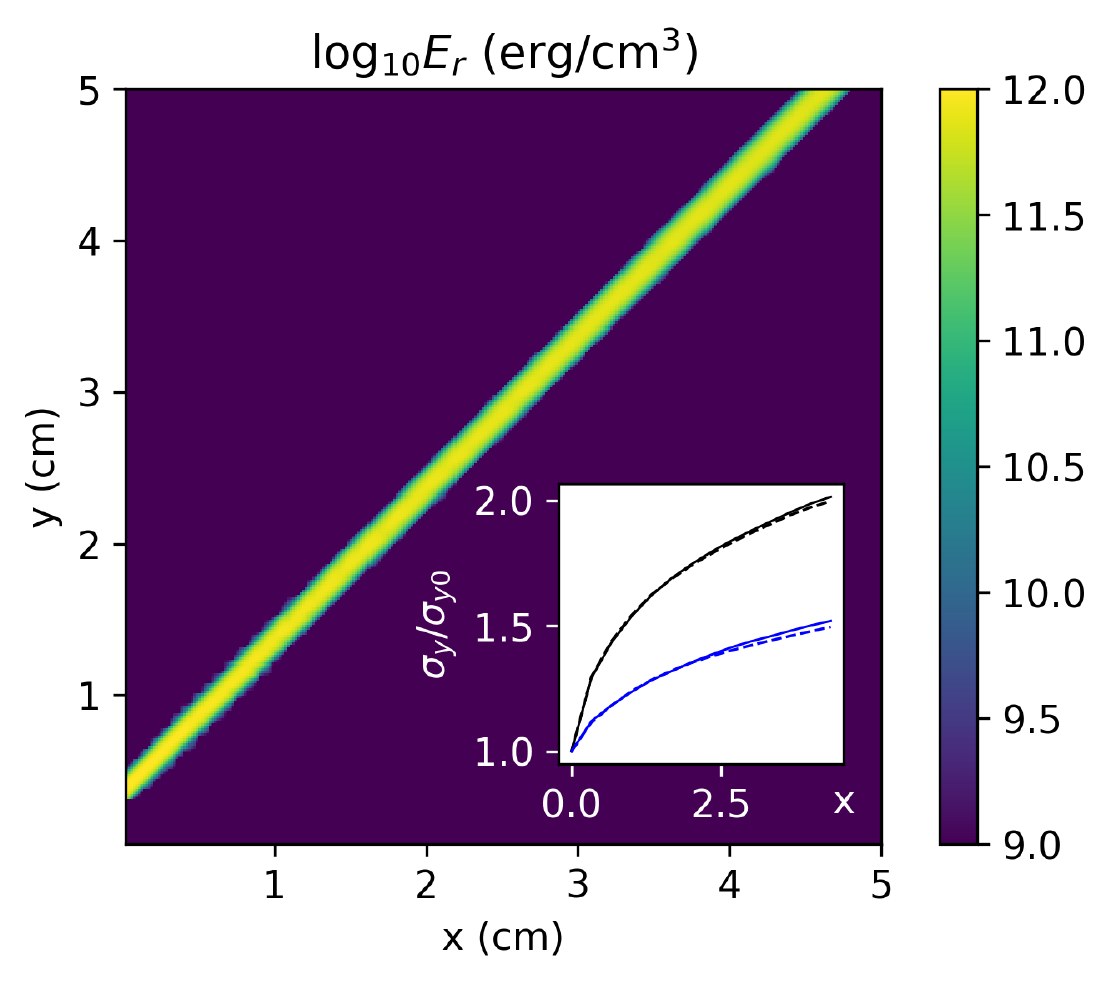}
	\caption{Free-streaming beam test.
     A radiation beam is introduced in a 2D grid
	 from its lower-left boundary, at $45\degree$
	 with respect to the
	 coordinate axes. The values of $\log_{10} E_r$ obtained with
	 the HLLC solver using a
	 resolution of $300\times300$ zones are plotted
	 as a function of $(x,y)$ at $t=5\times10^{-10}$ s (color scale).
	 The relative dispersion $\sigma_y/\sigma_{y0}$ along
	 the $y$ direction is shown in the
	 lower-right corner as a function of $x$ (cm), for the
	 selected resolutions of $150\times150$ (black lines)
	 and $300\times300$ (blue lines). In both cases, solid
	 and dashed lines correspond respectively to the results
	 obtained with the HLL and the HLLC solver.
	 % It is shown here
	 %that the beam's broadening is reduced
	 %when the resolution is increased.
	 }
	\label{fig:BeamTest}
  \end{figure}

 Again we compare the performance of the  HLL and HLLC
 solvers, using the fourth-order linear slopes
 by \cite{Miller2001} and resolutions of
 $150\times150$ and 
 $300\times300$ zones.
 The Courant number is $C_a=0.4$.
 The energy density distribution obtained with the HLLC solver
 at the largest resolution
 is shown in Fig. \ref{fig:BeamTest} at $t=5\times10^{-10}$ s.
 In every case, a beam forms and reaches
 the upper boundary between $x=4$ cm and $x=5$ cm,
 after crossing a distance equivalent to roughly $\sim 64$
 times its initial width.
 Since no interaction with matter is considered,
 photons should be transported in straight lines.
 As already mentioned, the free-streaming limit corresponds
 to a delta-like specific intensity parallel to $\mathbf{F}_r$.
 Hence, photons are injected in only one direction, and
 the beam's structure should be maintained as it crosses the
 computational domain.
 However, in the simulations, the beam broadens due to
 numerical diffusion before reaching the upper boundary.
 For this particular test, due to its strong discontinuities,
 we have seen that this effect is enhanced by the flattening
 applied during the reconstruction step
 in order to satisfy Equation \eqref{Eq:FsmallerE},
 which is necessary for stability reasons.
 
 In order to quantify this effect and its dependecy
 on the numerical resolution, we have computed several
 time-averaged $E_r(y)$ profiles along vertical
 cuts at different $x$ values.
 As an indicator of the beam's width, we have computed
 for each $x$
 the standard deviation of these profiles as
 \begin{equation}
 \sigma_y= \sqrt{\int_0^L \left[
 y - \overline{y} 
 \right]^2 \varphi(y)\, \mathrm{d}y}\,,
 \end{equation}
 with
 \begin{equation}
 \overline{y} =\int_0^L \varphi(y)\, y \,\mathrm{d}y \,,
 \end{equation}
% \begin{equation}
% \overline{y} =\left( \int_0^L E_r(y)\, y \,\mathrm{d}y \right)
% \Bigg{/} \left( \int_0^L E_r(y) \,\mathrm{d}y \right) .
% \end{equation}
%  \begin{equation}
% \overline{y} =\int_0^L E_r(y)\, y \,\mathrm{d}y 
% \bigg{/} \int_0^L E_r(y) \,\mathrm{d}y .
% \end{equation}
%  \begin{equation}
% \overline{y} = \frac{1}{\int_0^L E_r(y) \,\mathrm{d}y }
% \int_0^L E_r(y)\, y \,\mathrm{d}y \, .
% \end{equation}
 where the weighting function $f(y)$ is defined as
 \begin{equation}
 \varphi(y) = \overline{E}_r(y)\bigg/
 \int_0^L \overline{E}_r(y) \, \mathrm{d}y\,,
 \end{equation}
 being $\overline{E}_r$ the time-averaged value of $E_r$.
 We have then divided the resulting values of $\sigma_y$ by
 $\sigma_{y0}\equiv\sigma_y(x=0)$, in order to show the relative
 growth of the dispersion.
 The resulting values of $\sigma_y/\sigma_{y0}$ are shown in Fig.
 \ref{fig:BeamTest}, where it can be seen that the beam's dispersion
 grows with $x$.
 The difference between $\sigma_y/\sigma_{y0}$ and its ideal value
 ($\sigma_y/\sigma_{y0}\equiv 1$) gets reduced 
 by a factor between 2 and 2.5 when the highest resolution is used.
 In the same figure, it can be seen that the dispersion is only
 slightly reduced when the HLLC solver is used instead of HLL.
 A similar plot of $\sigma_y/\sigma_{y0}$ is obtained with the
 second-order limiter by \cite{vanLeer1974}, where the values
 of the relative dispersion increase roughly between $30\%$ and $40\%$,
 showing as in Section \ref{S:ShockThin} that the accuracy of these
 methods not only depends on the chosen Riemann solver but it is also
 extremely sensitive to the chosen reconstruction scheme.

%%%%%%%%%%%%%%%%%%%%%%%%%%%%%%%%%%%%%%%%%%%%%%%%%%%%%%%%%%%%%%%%%%%%%%%%
 \subsection{Radiation-matter coupling}\label{S:RadMatCoup}
%%%%%%%%%%%%%%%%%%%%%%%%%%%%%%%%%%%%%%%%%%%%%%%%%%%%%%%%%%%%%%%%%%%%%%%%

 In order to verify the correct integration
 of the interaction terms, we have
 run a test proposed by \citet{TurnerStone2001},
 in which matter and radiation approach thermal equilibrium
 in a homogeneous system. This is achieved by solving the
 Rad-RHD equations in a single-cell grid, thus removing any
 spatial dependence.
 In this configuration, due to the form of Equations
 \eqref{Eq:RadRMHD}-\eqref{Eq:RadRMHD2}, all the fields but
 the energy densities of both radiation and matter remain constant 
 for $t>0$.
 Using conservation of total energy, the resulting equation for
 the evolution of the gas energy density  (in cgs units) 
 is 
 \begin{equation}\label{Eq:RadMatCoup}
 \frac{1}{c}\frac{\partial \mathcal{E}}{\partial t} = \rho \kappa \left(
 E_r - 4\pi B\left( T \right)
 \right).
 \end{equation} 
 This can be simplified if the chosen initial conditions are
 such that $E_r$ is constant throughout the system's evolution. In
 that case, Equation \eqref{Eq:RadMatCoup} can be solved
 analytically, leading to an implicit relation between $\mathcal{E}$
 and $t$ that can be inverted using standard methods.
 
 We have run this test for two different initial conditions,
 using in both cases $\rho=10^{-7}$ \mbox{g cm$^{-3}$},
 $E_r= 10^{12}$ \mbox{erg cm$^{-3}$}, opacities $\kappa=0.4$
 \mbox{cm$^{2}$ g$^{-1}$} and $\sigma = 0$,
 and a mean molecular weight $\mu=0.6$.
 A constant-gamma EoS has been assumed, with $\Gamma=5/3$.
 We have chosen the initial gas energy density
 to be either $\mathcal{E}=10^{10}$
 \mbox{erg cm$^{-3}$} or $\mathcal{E}=10^2$ \mbox{erg cm$^{-3}$},
 which are, respectively, above and below the final equilibrium
 value, of around $7\times 10^7$ \mbox{erg cm$^{-3}$}. 
  
 \begin{figure}[t]
	\centering
	\includegraphics[width=0.47\textwidth]{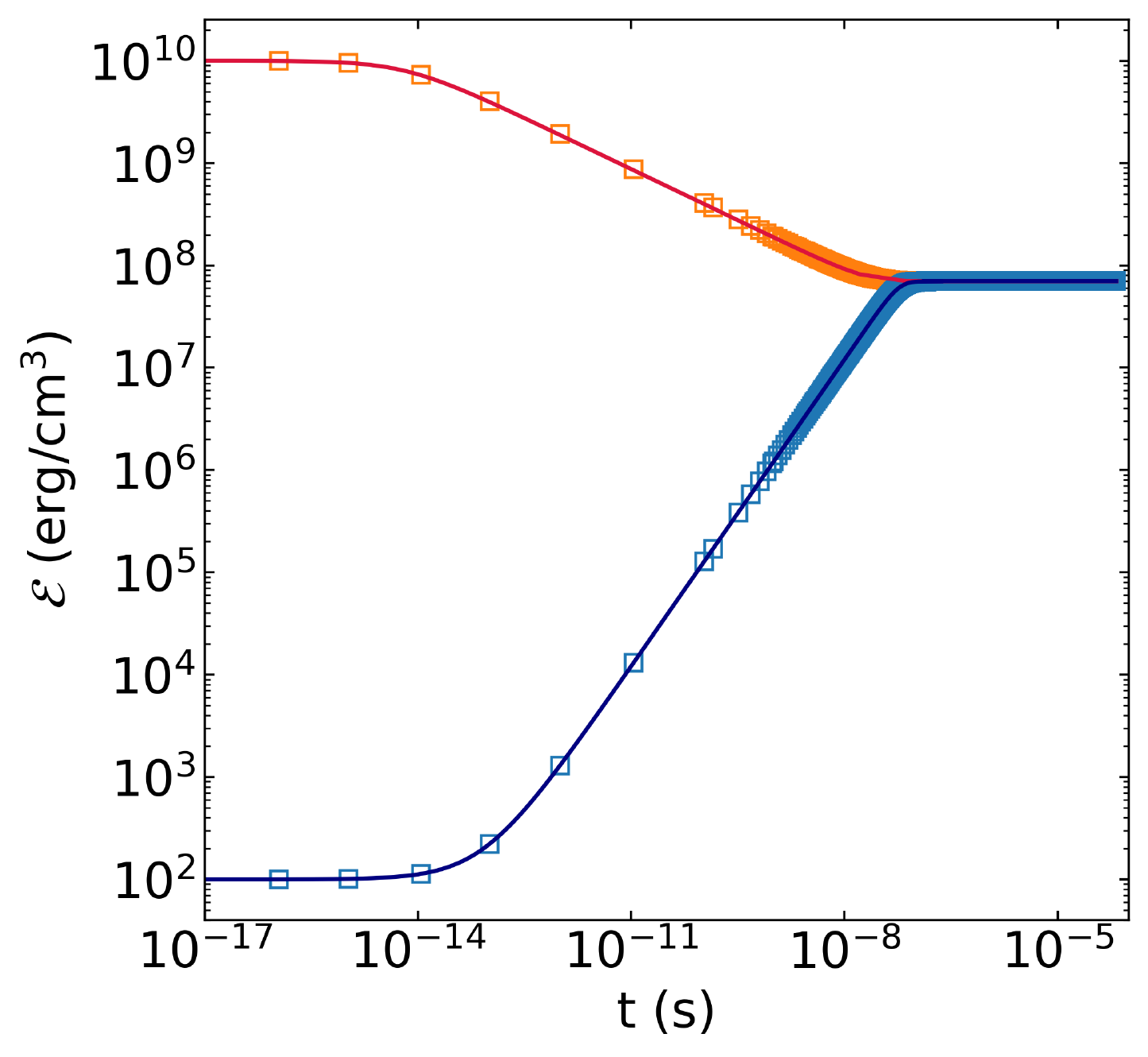}
	\caption{Radiation-matter coupling test. The gas energy density
	$\mathcal{E}$ is plotted as a function of time
	for the two chosen initial conditions,
	until thermal equilibrium is reached.
	The obtained numerical values (empty squares) 
	are shown here to match the analytical solutions (solid lines)
	for both initial conditions.}
	\label{fig:CoupRadMatter}
  \end{figure}
  
 The gas energy density is plotted as a function
 of time for both conditions in Fig. \ref{fig:CoupRadMatter}.
 Simulations are started from $t=10^{-10}$ s, with an initial
 time step $\Delta t = 10^{-10}$ s. An additional run between
 $t=10^{-16}$ s and $10^{-10}$ s is done for each initial
 condition with an initial $\Delta t = 10^{-16}$ s,
 in order to show the evolution in the initial stage.
 In every case, the gas radiation energy goes through an initial
 constant phase that lasts until $t\sim 10^{-14}$ s, after which
 it varies towards the equilibrium value. Equilibrium
 occurs when the condition $E_r=4\pi B(T)$ is reached
 (see Eq. \eqref{Eq:RadMatCoup}), i.e.,
 when the power emitted by the gas equals its energy absorption
 rate.
 This happens around $t\approx 10^{-7}$ s
 for both initial conditions.
 As shown in Fig. \ref{fig:CoupRadMatter}, the numerical solutions
 match the analytical ones in the considered time range.
 
%%%%%%%%%%%%%%%%%%%%%%%%%%%%%%%%%%%%%%%%%%%%%%%%%%%%%%%%%%%%%%%%%%%%%%%%
\subsection{Shock waves}\label{S:Shocks}
%%%%%%%%%%%%%%%%%%%%%%%%%%%%%%%%%%%%%%%%%%%%%%%%%%%%%%%%%%%%%%%%%%%%%%%%

%\startlongtable
 \begin{deluxetable*}{c|cccccccccc} 
 \tablecaption{Parameters used in the shock tests, in code units.
	The subscripts $R$ and $L$ correspond, respectively, to the 
	initial conditions for $x>0$ and $x<0$. \label{Table:ShockParams} }
	\tablewidth{0pt}
 \tablehead{
	\colhead{Test} & \colhead{$\rho_L$} & \colhead{$p_{g,L}$} &
	\colhead{$u^x_L$}  &\colhead{$\tilde{E}_{r,L}$} 
	& \colhead{$\rho_R$} & \colhead{$p_{g,R}$} &
	\colhead{$u^x_R$}  &\colhead{$\tilde{E}_{r,R}$}
	&\colhead{$\Gamma$} &\colhead{$\kappa$} 
	}
%\colnumbers
\startdata
    1 & $1.0$ & $3.0\times 10^{-5}$ & $0.015$ & $1.0\times 10^{-8}$ &
    $2.4$ & $1.61\times 10^{-4}$ & $6.25\times 10^{-3}$
    & $2.51\times 10^{-7}$ & $5/3$ & $0.4$ \\ 
    2 & $1.0$ & $4.0\times 10^{-3}$ & $0.25$ & $2.0\times 10^{-5}$ &
    $3.11$ & $0.04512$ & $0.0804$
    & $3.46\times 10^{-3}$ & $5/3$ & $0.2$ \\ 
    3 &$1.0$ & $60.0$ & $10.0$ & $2.0$ &
    $8.0$ & $2.34\times 10^{3}$ & $1.25$
    & $1.14\times 10^{3}$ & $2$ & $0.3$ \\ 
    4 & $1.0$ & $6.0\times 10^{-3}$ & $0.69$ & $0.18$ &
    $3.65$ & $3.59\times 10^{-2}$ & $0.189$
    & $1.3$ & $5/3$ & $0.08$ \\
\enddata
 \end{deluxetable*}

 We now study the code's ability to reproduce general shock-like
 solutions without neglecting the interaction terms.
 To this purpose, we have reproduced a series of tests
 proposed by \cite{Farris2008}.
 As in Section \ref{S:ShockThin}, we place a single initial
 discontinuity at the center of the one-dimensional domain
 defined by the interval \mbox{$[-20,20]$}.
 At $t=0$, both matter and radiation fields are constant on
 each side  of the domain, and satisfy the condition
 for LTE between matter and radiation, that is,
 \mbox{$\tilde{E}_r=4\pi B(T)$}.
 Additionally, the fluxes on each side obey 
 $\tilde{F}_r^x = 0.01 \times \tilde{E}_r$.
 A constant-gamma EoS is assumed, scattering
 opacity is neglected, and a Courant factor \mbox{$C_a=0.25$}
 is used. 
  
 Initial conditions are chosen in such a way that the system evolves
 until it reaches a final stationary state.
 Neglecting time derivatives,
 Equations \eqref{Eq:RadRMHD}-\eqref{Eq:RadRMHD2} lead to 
 \begin{align} \label{Eq:Shock1}
   \partial_x \left(\rho u^x\right) &= 0 \\ \label{Eq:Shock2}
   \partial_x \left(m_{tot}^x\right) &= 0 \\ \label{Eq:Shock3}
   \partial_x \left( m^x v^x + p_g + P_r^{xx} \right) &= 0 \\
  \label{Eq:Shock4}
   \partial_x \left(F_r^x \right)&= -G^0 \\ \label{Eq:Shock5}
   \partial_x \left(P_r^{xx}\right) &= -G^x. 
 \end{align}
 A time-independent solution demands that quantities under derivative
 in Equations \eqref{Eq:Shock1}--\eqref{Eq:Shock3} remain constant,
 and this condition must also be respected by the initial states. 
 In addition, Equations \eqref{Eq:Shock4} to \eqref{Eq:Shock5} show
 that the final $F_r^x$ and $P_r^{xx}$ must be continuous,
 although their derivatives can be discontinuous.
 This does not necessarily imply that the final $E_r$
 profile must also be continuous, since any value
 of $P_r^{xx}(E_r,F^x_r)$ can correspond to up to two
 different $E_r$ values for fixed $F^x_r$.
 However, in the particular case where \mbox{$F^x_r<P_r^{xx}$},
 it can be shown using Eqs. \eqref{Eq:M11}-\eqref{Eq:M13}
 that the inversion of $P_r^{xx}(E_r,F^x_r)$
 in terms of $E_r$ leads to unique solutions,
 and thus $E_r$ must be continuous.
 In the same way, we have verified that this condition
 is equivalent to \mbox{$F^x_r/E_r<3/7$}.
 
  We have performed four tests for different physical regimes.
  %For each of these, we shall indicate with the subscripts
  %$R$ and $L$
  %the initial conditions for $x>0$ and $x<0$, respectively.
  All the initial values are chosen to coincide with those in 
  \citet{Farris2008}. In that work, as in several others where the
  same tests are performed \citep[see e.g.][]{Zanotti2011,
  Fragile2012,Sadowski2013}, the Eddington approximation, given by
  \mbox{$\tilde{P}^{xx}_r=\tilde{E}_r/3$}, is used instead of the M1
  closure. Therefore, our results are not comparable with these unless
  the final state satisfies \mbox{$\tilde{P}^{xx}_r\simeq\tilde{E}_r/3$}
  in the whole domain. We now outline the main features of each test,
  whose parameters are summarized in Table \ref{Table:ShockParams}:
  
  \begin{figure}[t]
	\centering
	\includegraphics[width=0.47\textwidth]{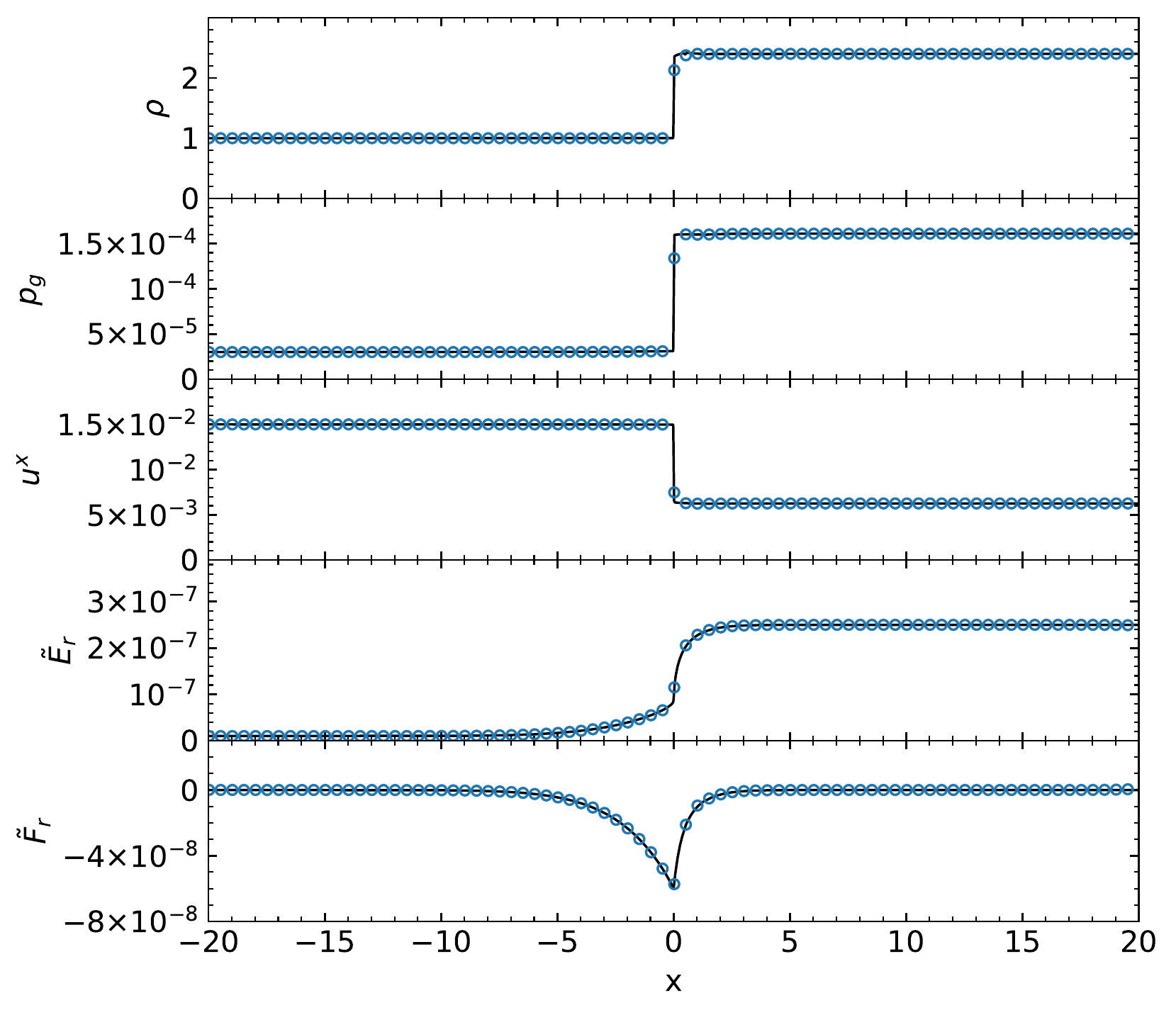}
	\caption{Final profiles of the nonrelativistic
	strong shock test, obtained using 3200 zones (solid black line)
	and 800 zones (empty blue circles, plotted every 10 values). }
	\label{fig:ShockTest1}
  \end{figure}
  
  \begin{figure}[t]
	\centering
	\includegraphics[width=0.47\textwidth]{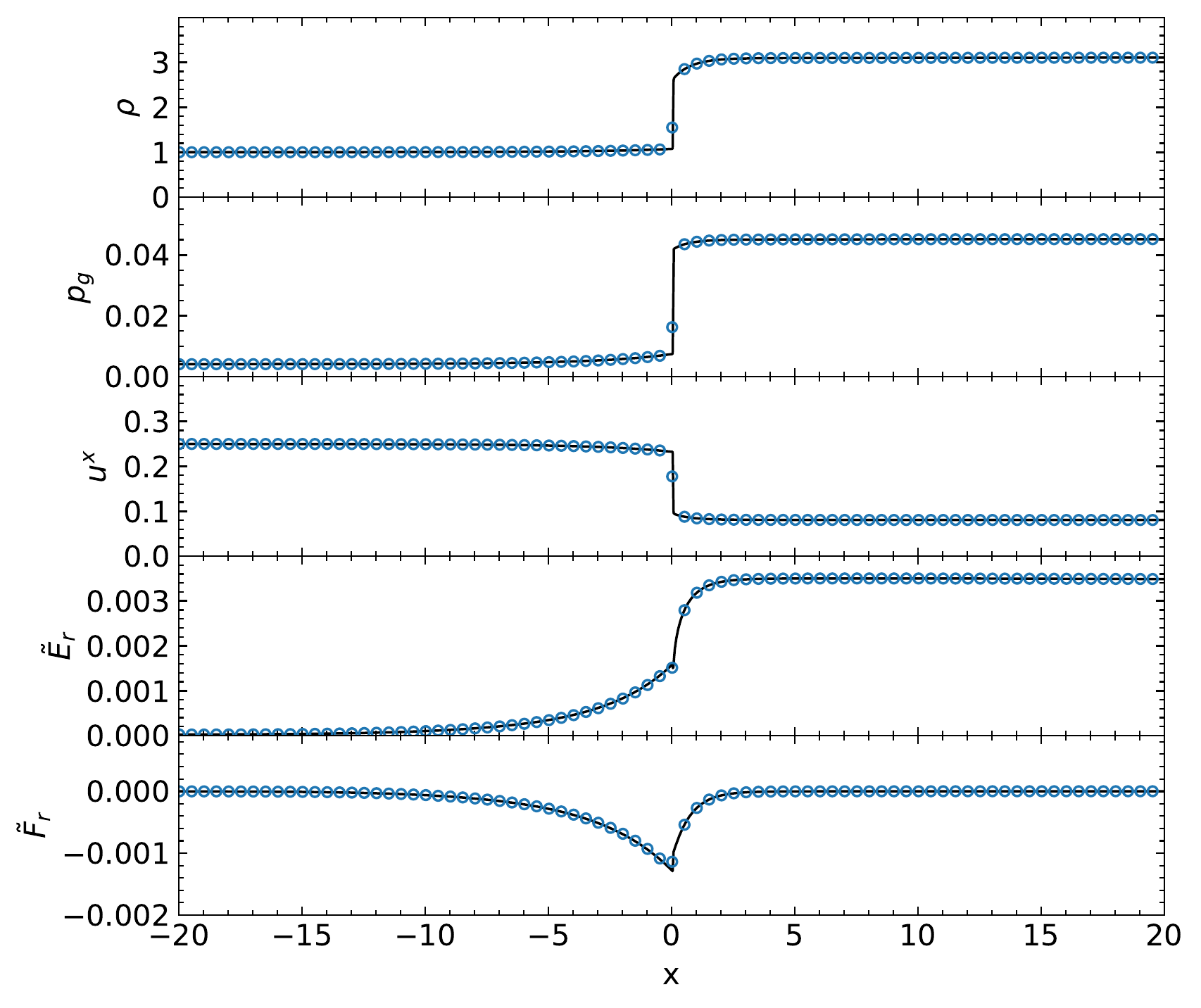}
	\caption{Same as Fig. \ref{fig:ShockTest1},
	for the mildly relativistic shock test.}
	\label{fig:ShockTest2}
  \end{figure}    
  
\begin{figure}[t]
	\centering
	\includegraphics[width=0.47\textwidth]{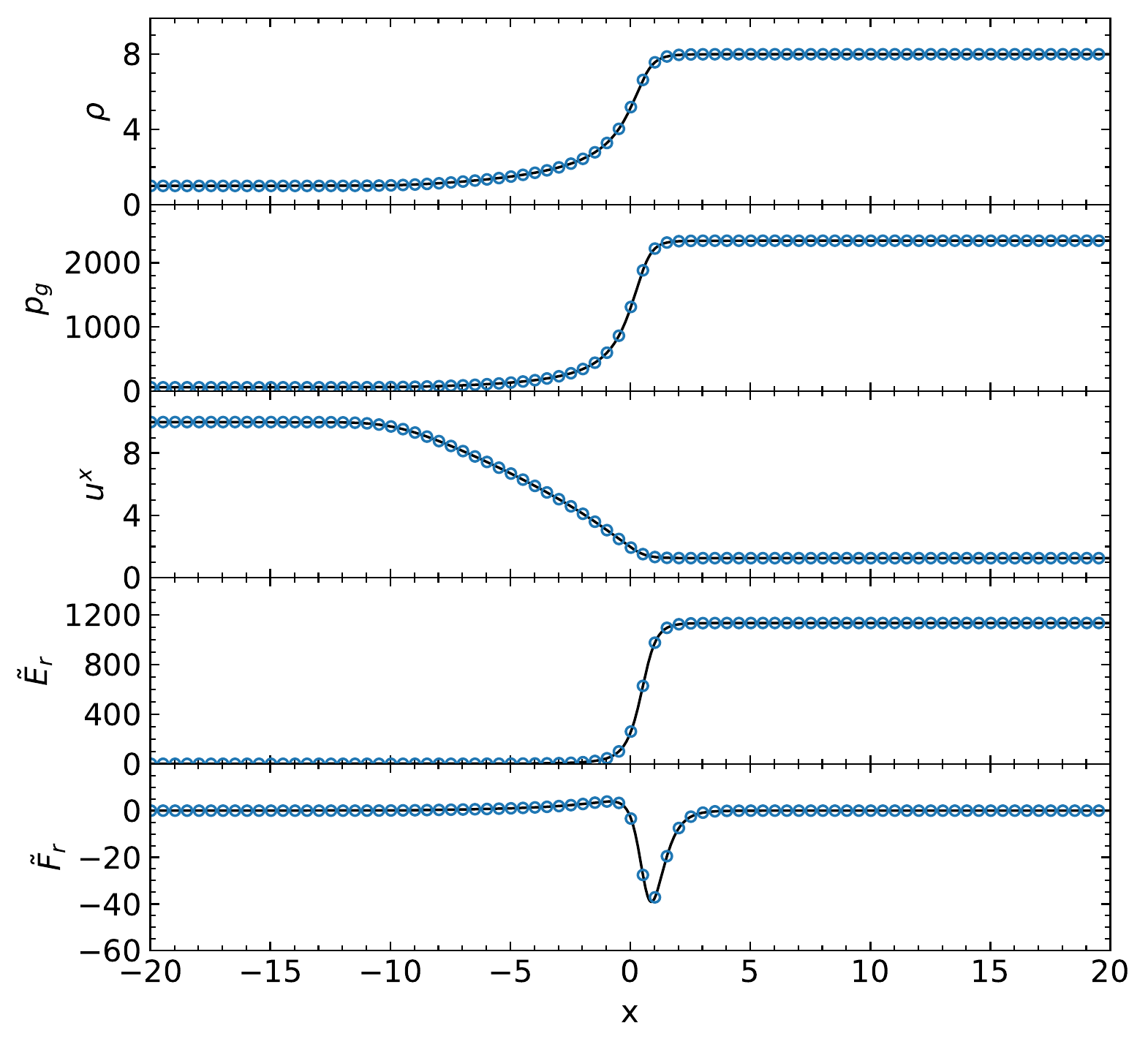}
	\caption{Same as Fig. \ref{fig:ShockTest1},
	for the highly relativistic wave test.}
	\label{fig:ShockTest3}
  \end{figure}

\begin{figure}[t]
	\centering
	\includegraphics[width=0.47\textwidth]{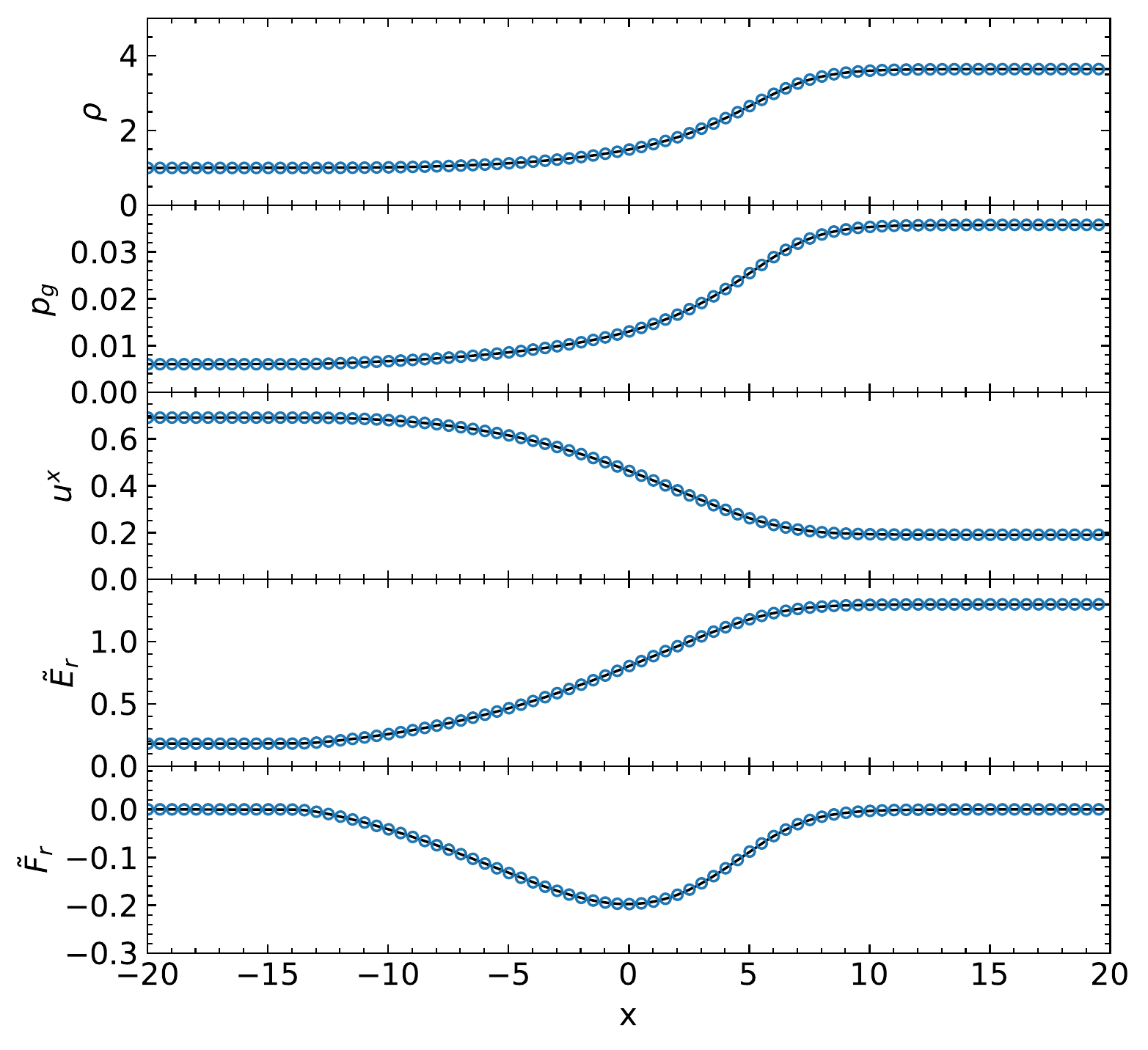}
	\caption{Same as Fig. \ref{fig:ShockTest1},
	for the radiation-pressure-dominated wave test.}
	\label{fig:ShockTest4}
  \end{figure}

    \begin{enumerate}
  \item \emph{Nonrelativistic strong shock}.
  A gas-pressure-dominated shock moves at a nonrelativistic speed
  in a cold gas (\mbox{$p_g\ll \rho$}),
  with a maximum $u^x$ of 0.015.
  The final profiles of $\rho$, $p_g$, $u^x$, $\tilde{E}_r$,
  and  $\tilde{F}^x_r$ are shown in Fig. \ref{fig:ShockTest1}. As in the
  non-radiative case, the first three show an abrupt change at $x=0$,
  while radiation fields seem continuous.
  
  \item \emph{Mildly relativistic strong shock}.
  The conditions are similar to the previous test, with the
  difference that a mildly relativistic velocity ($u^x\le 0.25$)
  is chosen. The final profiles (see Fig. \ref{fig:ShockTest2})
  look similar to those in Fig. \ref{fig:ShockTest1}, with the
  difference that $\tilde{E}_r$ exhibits a small discontinuity close to
  $x=0$.
  
  \item \emph{Highly relativistic wave}.
  Initial conditions are those of a highly relativistic
  gas-pressure-dominated wave (\mbox{$u^x\le 10$},
  \mbox{$\rho\ll\tilde{P}^{xx}_r < p_g$}).
  In this case, as it can be seen in Fig. \ref{fig:ShockTest3},
  all the profiles are continuous.
  
  \item \emph{Radiation-pressure-dominated wave}. 
  In this case we study a situation where the radiation pressure
  is much higher than the gas pressure, in a shock that propagates
  at a mildly relativistic velocity ($u^x\le 0.69$). As in the
  previous case, there are no discontinuities in the final profiles
  (see Fig. \ref{fig:ShockTest4}).
   \end{enumerate}
   
  In order to test the convergence of the numerical solutions, we
  have performed each simulation twice, dividing the domain in $800$
  and in $3200$ zones. In every case, as shown in Figs.
  \ref{fig:ShockTest1}-\ref{fig:ShockTest4}, both solutions
  coincide. However, our results do not coincide with those
  obtained in the references mentioned above.
  The most noticeable case is the test shown in Fig. \ref{fig:ShockTest2},
  where the ratio $\tilde{P}^{xx}_r/\tilde{E}_r$
  reaches a maximum value of $0.74$ close to
  the shock, instead of the value of $1/3$ that would be obtained
  within the Eddington approximation.
  The result is a much smoother $\tilde{E}_r$ profile than
  the one shown in, for instance, \citet{Farris2008}. Yet,
  our results show a good agreement with those in
  \citet{Takahashi2013}, where the tests are also performed
  assuming the M1 closure.

	We point out that, in the nonrelativistic strong shock case,
    characteristic fluid speeds are $\sim 35$
	times smaller than those corresponding to radiation transport.
	Still, computations
	do not show significant increase of numerical diffusion owing
	to such scale disparity. The
	same conclusion holds if computations are done in the
	downstream reference frame (not shown here).

%%%%%%%%%%%%%%%%%%%%%%%%%%%%%%%%%%%%%%%%%%%%%%%%%%%%%%%%%%%%%%%%%%%%%%%%
 \subsection{Radiation pulse}\label{S:RadPulse}
%%%%%%%%%%%%%%%%%%%%%%%%%%%%%%%%%%%%%%%%%%%%%%%%%%%%%%%%%%%%%%%%%%%%%%%%

 Following \citet{Sadowski2013}, we have tested the evolution of
 a radiation pulse in the optically thin and optically thick limits.
 These two regimes allowed us to assess, respectively, the code performance when
 choosing different coordinate systems and its accuracy in
 the diffusion limit, as summarized below.

%%%%%%%%%%%%%%%%%%%%%%%%%%%%%%%%%%%%%%%%%%%%%%%%%%%%%%%%%%%%%%%%%%%%%%%%
 \subsubsection{Optically thin case}\label{S:PulseThin}
%%%%%%%%%%%%%%%%%%%%%%%%%%%%%%%%%%%%%%%%%%%%%%%%%%%%%%%%%%%%%%%%%%%%%%%%
 
 We considered an initial spherically symmetric radiation energy
 distribution, contained around the center of a 3D box of side
 \mbox{$L=100$}.
 Radiation energy is initially set as $E_r=4\pi B(T_r)$, with
 \begin{equation}
 T_{r}= T_0\left(1+100\, e^{-r^2/w^2}\right),
 \end{equation}
 where $r$ is the spherical radius, while
 
 $T_0=10^6$ and $w=5$. Similarly, gas pressure is
 set in such a way that $T(\rho,p_g)=T_0$, which means that
 the system is initially in thermal equilibrium far from the
 pulse. 
 We also set $\rho=1$, $v^x=0$ and $F_r^x=0$ in the whole domain,
 $\Gamma=5/3$, $C_a=0.4$,
 $\kappa=0$, and a small scattering
 opacity $\sigma=10^{-6}$. In this way, the total optical depth 
 from side to side of the box is
 \mbox{$\tau=\rho\,\sigma L=10^{-4}\ll 1$}, i.e., the box is
 transparent to radiation.

 We have computed the departure from these conditions using
 1D spherical and 3D Cartesian coordinates.
 In the Cartesian case, we have employed a uniform grid resolution
 of \mbox{$200\times200\times200$} zones. 
 On the other hand, in spherical geometry, 
 our domain is the region $r\in[0,L/2]$ 
 using a uniformly spaced grid of $100$ zones, in order to have a
 comparable resolution with the 3D simulations.
 In this last case, reflective boundary conditions have
 been set at $r=0$.
 
\begin{figure}[t]
	\centering
	\includegraphics[width=0.47\textwidth]{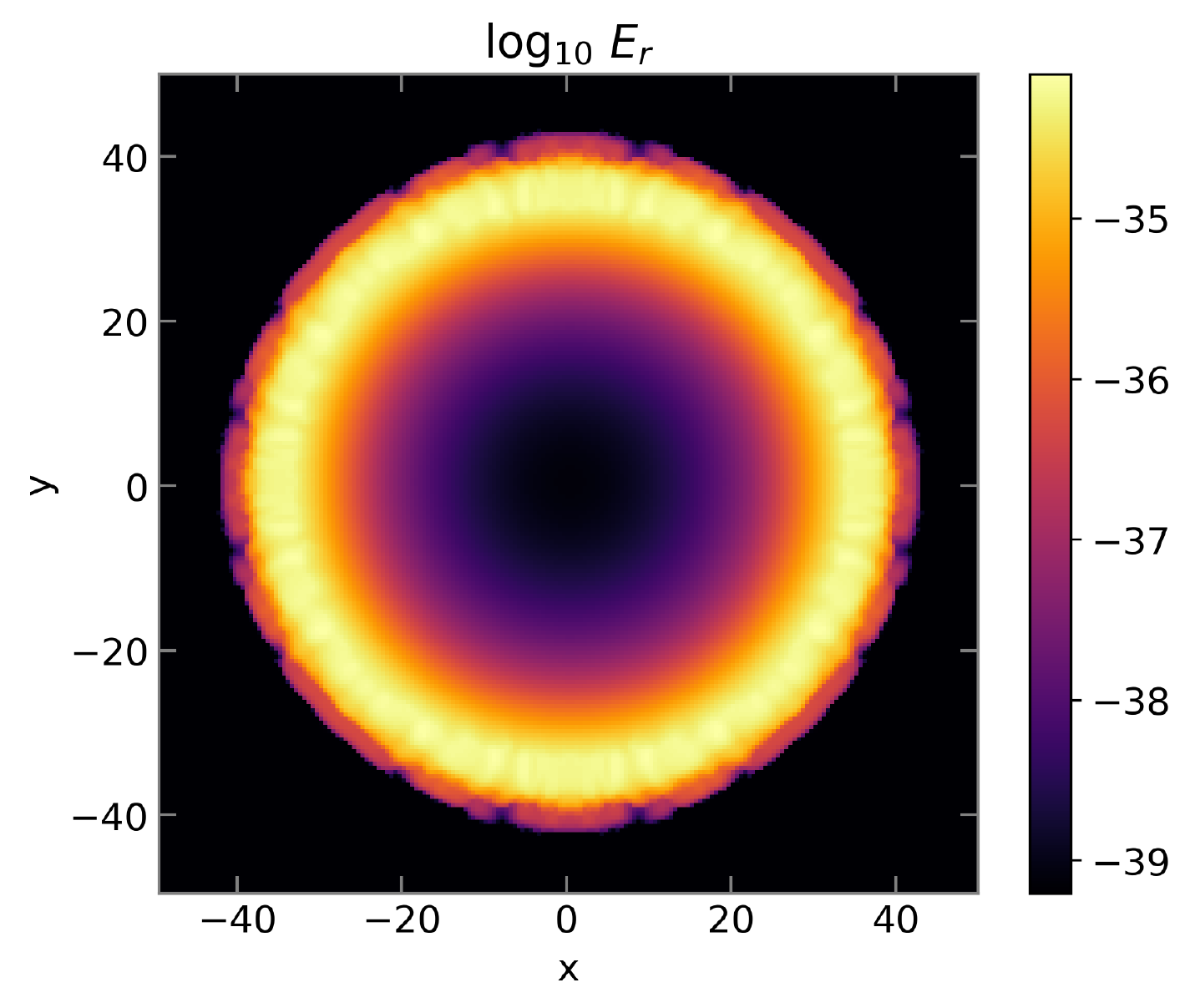}
	\caption{ Radiation energy density map of the optically thin
	radiation pulse computed using a \mbox{$200\times200\times200$}
	uniform Cartesian grid.	 Values of $\log_{10}E_r$ on the plane
	$z=0$ are shown at $t=35$, when the blast wave has already
	been formed.}
	\label{fig:PulseThin1}
  \end{figure}

\begin{figure}[t]
	\centering
	\includegraphics[width=0.47\textwidth]{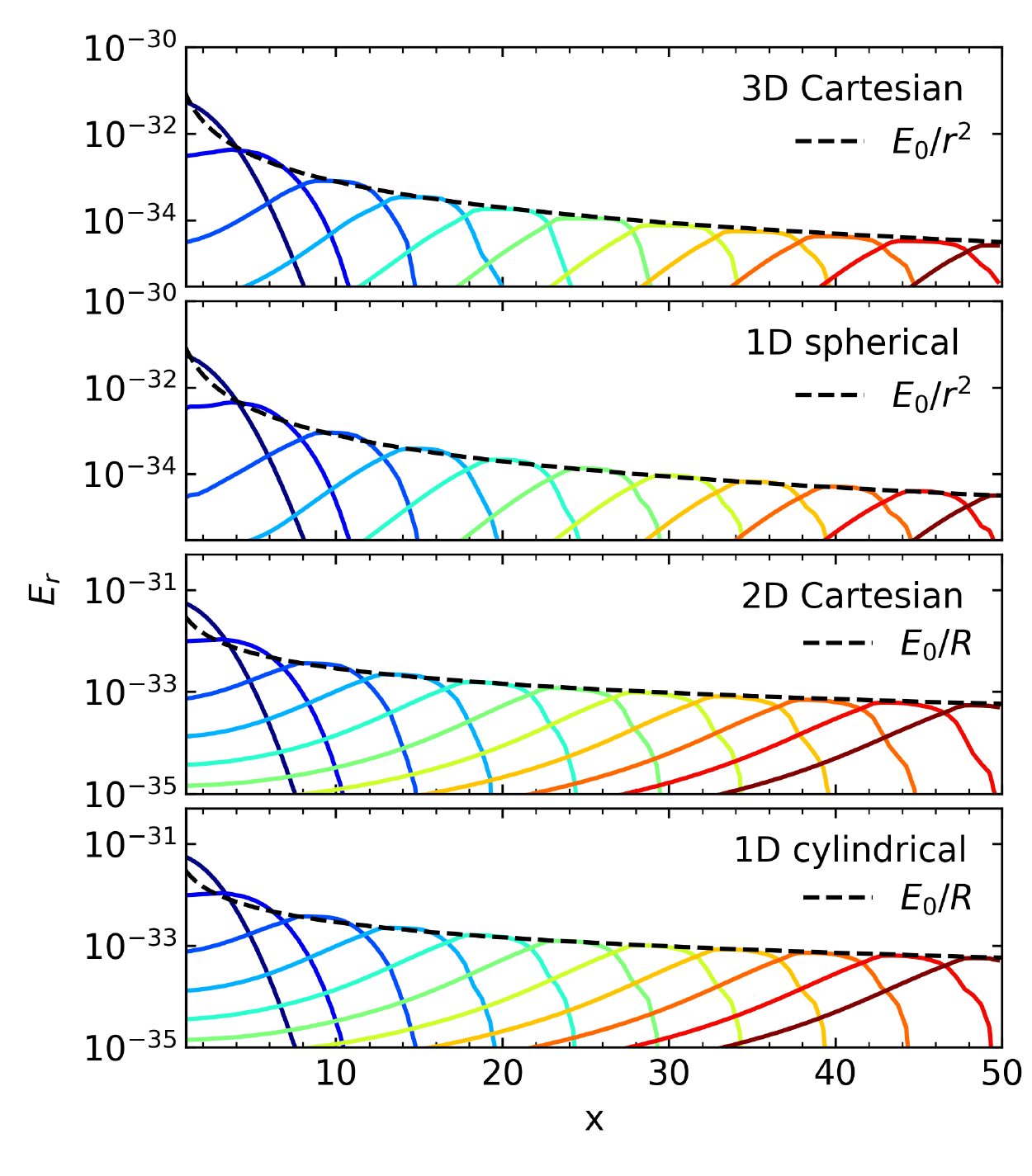}
	\caption{Radiation energy density profiles in the optically
	thin pulse test	(solid colored lines),	computed at $y=z=0$.
	From above: profiles obtained
	using 3D Cartesian,
	1D spherical, 2D Cartesian and 1D cylindrical coordinates,
	shown every $\Delta t = 5.0$. The dependence of the
	maximum energy on $1/r^2$ ($1/R$) is shown with dashed black
	lines in the first (last) two cases.}
	\label{fig:PulseProfiles}
  \end{figure}

 As shown in Fig. \ref{fig:PulseThin1}, the pulse expands and forms
 a nearly isotropic blast wave, which slightly deviates from the
 spherical shape in the Cartesian case due to grid noise.
 The evolution of the radiation energy profiles in both simulations
 is shown in the two upper panels of Figure \ref{fig:PulseProfiles}.
 Since no absorption in the material is considered, the total
 radiation energy is conserved, and thus the maximum energy density
 of the formed expanding wave decreases as \mbox{$1/r^2$}.
 As it can be seen in Fig. \ref{fig:PulseProfiles}, this
 dependence is effectively verified once the
 blast wave is formed. The same kind of analysis is possible if
 radiation is contained entirely on the plane \mbox{$z=0$}.
 In this case, the maximum energy density decreases
 as \mbox{$1/R$}, with \mbox{$R=\sqrt{x^2+y^2}$}.
 We have verified this behavior in 1D
 cylindrical and 2D Cartesian coordinates,
 employing uniform grids of $100$ zones
 in the first case and 	\mbox{$200\times200$} in the second
 (see the two lower panels in Fig. \ref{fig:PulseProfiles}).
 In every case, the same simulations performed with different
 coordinates systems show a good agreement.

%%%%%%%%%%%%%%%%%%%%%%%%%%%%%%%%%%%%%%%%%%%%%%%%%%%%%%%%%%%%%%%%%%%%%%%%
 \subsubsection{Optically thick case}\label{S:PulseThick}
%%%%%%%%%%%%%%%%%%%%%%%%%%%%%%%%%%%%%%%%%%%%%%%%%%%%%%%%%%%%%%%%%%%%%%%%

 We now consider the case where the scattering opacity is nine
 orders of magnitude larger than in the previous simulations, i.e.,
 \mbox{$\sigma=10^3$}, and all the other parameters remain unchanged.
 In that situation, the optical thickness from side to side of the
 box is \mbox{$\tau = 10^5\gg 1$}, which means that the box is 
 largely opaque to radiation.
 Here we solve the evolution equations on a
 Cartesian one-dimensional grid
 with uniform spacing.
 Using a resolution of 101 zones, the
 optical thickness of a single cell is $\tau\sim10^3$. For this
 reason, signal speeds are always limited accordingly to Eq.
 \eqref{Eq:RadSpeedLim}.
 
\begin{figure}[t]
	\centering
	\includegraphics[width=0.47\textwidth]{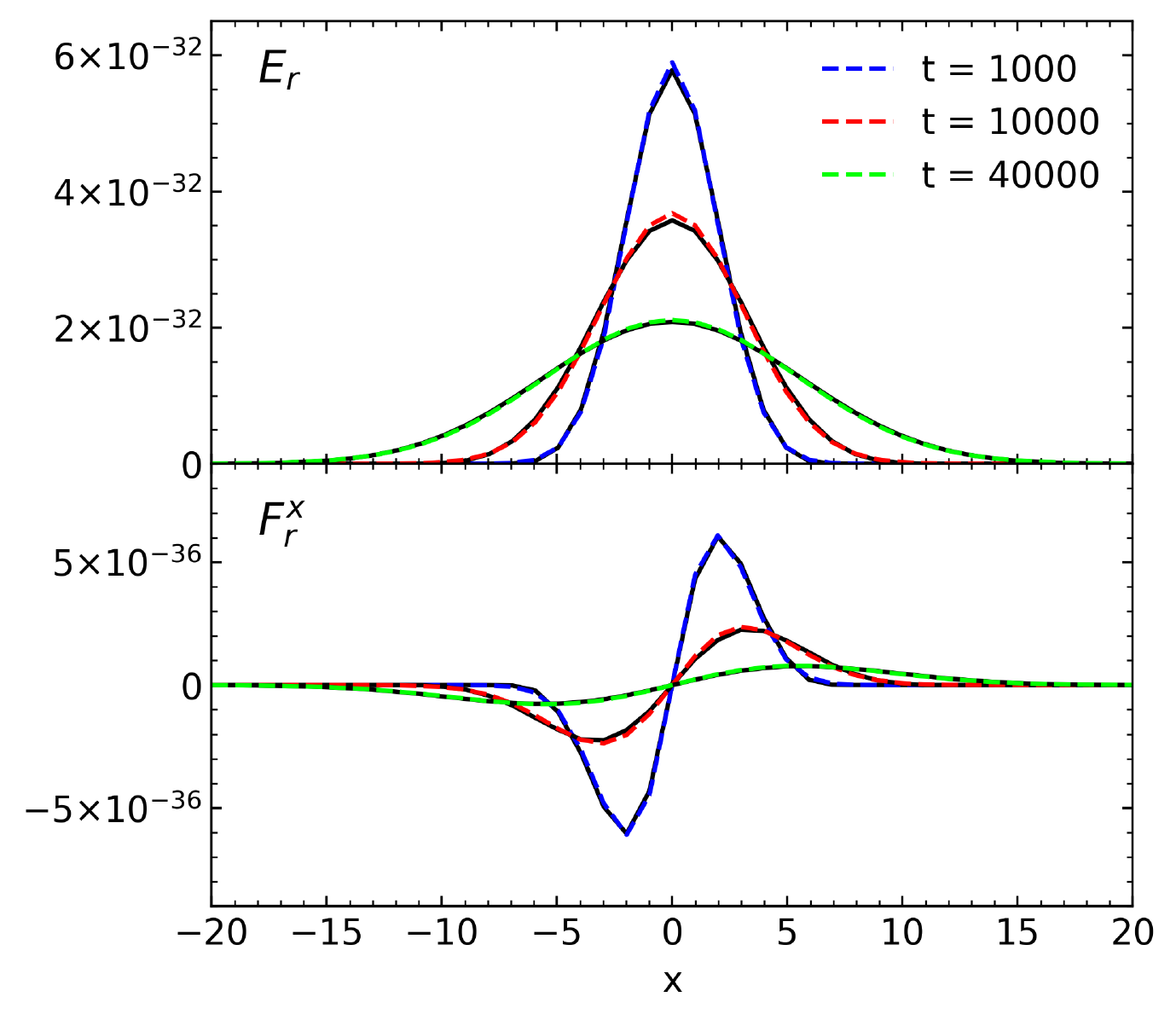}
	\caption{Radiation energy density and flux profiles
	in the optically thick pulse test, shown at $t=10^3, 10^4$
	and $4\times10^4$ (solid black lines). The analytical solution of
	the diffusion equation (Eq. \eqref{Eq:DiffEq}) is superimposed
	(dashed colored lines).
	}
	\label{fig:DiffEq}
  \end{figure}

 Under these conditions, the system evolves in such a way that
 \mbox{$\vert\partial_t F_r^x\vert\ll \vert \partial_x P_r^{xx}
 \vert$} and \mbox{$\vert F_r^x\vert\ll E_r$}, and therefore
 \mbox{$P_r^{xx}\simeq E_r/3$}, as pointed out in Section
 \ref{S:M1}. Neglecting the term \mbox{$\partial_t F_r^x$} in Eq.
 (\ref{Eq:RadRMHD2}) and assuming $P_r^{xx}=E_r/3$, the radiation
 flux can be written as \mbox{$F_r^x=-\partial_x E_r /3\rho \chi $}.
 Hence, assuming the density to remain constant,
 the radiation energy density
 should evolve accordingly to the following diffusion equation:
 \begin{equation}\label{Eq:DiffEq}
 \frac{\partial E_r}{\partial t} = \frac{1}{3\rho \chi}
 \frac{\partial^2 E_r}{\partial x^2} .
 \end{equation}
 With the chosen initial conditions, this equation can be solved
 analitically, e.g., by means of a Fourier transform
 in the spatial domain.
 The exact and numerical solution are shown in Fig.
 \ref{fig:DiffEq}.
 Our results show a good agreement between the analytical
 and numerical solutions. Furthermore, we have verified that, if
 radiation-matter interaction is not taken into account for the
 signal speed calculation, i.e., if the limiting given by Eq.
 \eqref{Eq:RadSpeedLim} is not applied, the pulse gets damped much
 faster than what it should be expected from Eq. \eqref{Eq:DiffEq},
 due to the numerical diffusion that occurs when signal speeds
 are overestimated.
 
 We have observed that this test leads
 to inaccurate values of $F_r^x$ if IMEX-SSP2(2,2,2) is used, although
 the values of $E_r$ remain close to the analytical ones.
 This problem lies in the fact that both the gradient of the flux of
 $F^x_r$ and its source term largely exceed
 $F^x_r$ and are not compensated in the last
 explicit step of the method (see Eq. \eqref{Eq:IMEX1}).
 When these conditions are met, we have observed that
 IMEX-SSP2(2,2,2) can lead to inaccuracies and instabilities due to
 failure in preserving energy positivity
 (see Section \ref{S:Shadows}).
 On the contrary, IMEX1 shows better performances in those cases,
 as flux and source terms are more accurately balanced during
 the implicit steps (see Eq. \eqref{Eq:IMEX2}).

\begin{figure}[t]
	\centering
	\includegraphics[width=0.47\textwidth]{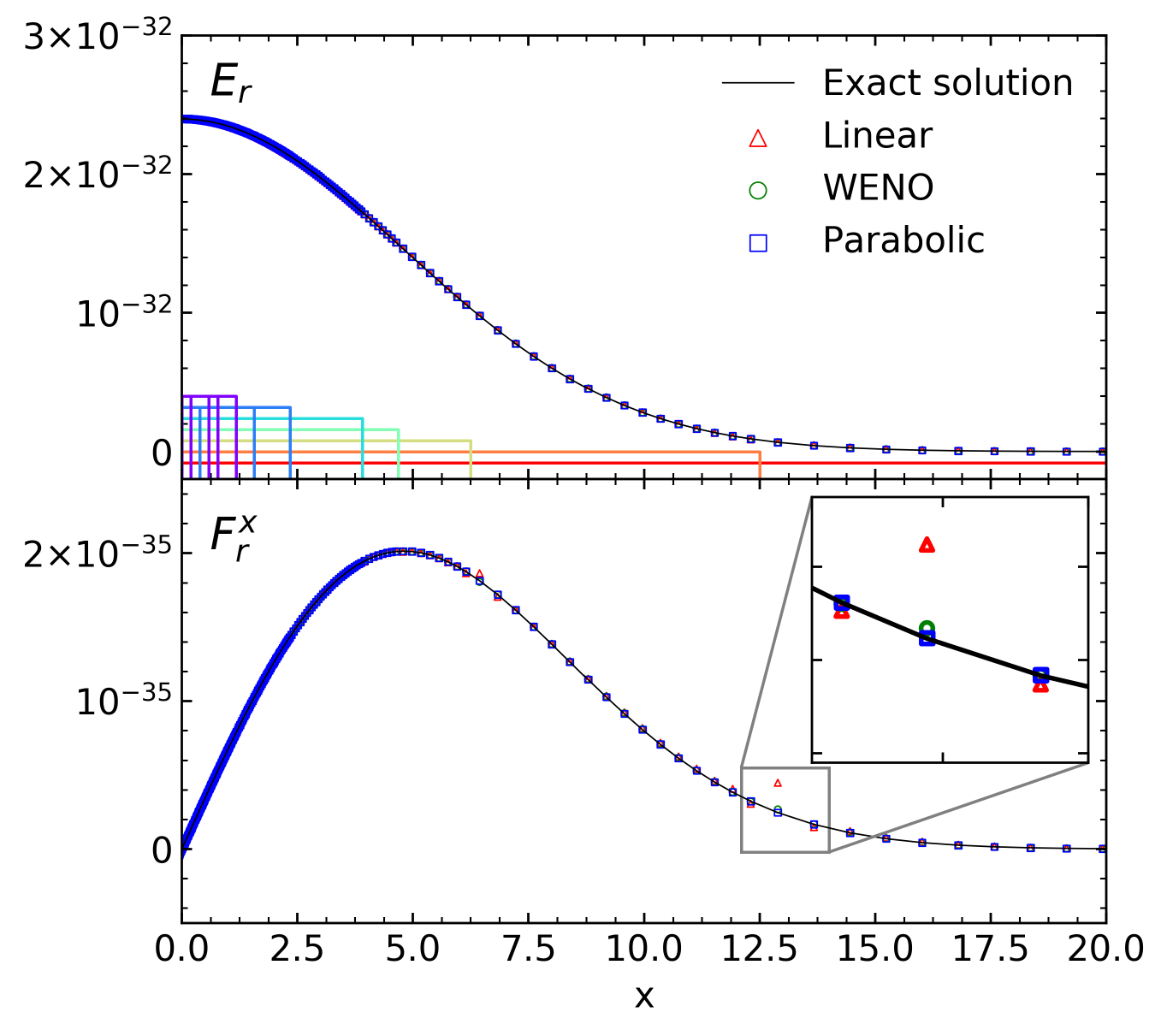}
	\caption{ Radiation energy density (top) and flux (bottom) profiles
	in the optically thick pulse test with $\sigma=50$ using a static
        AMR grid with six refinement levels.
        Solution are shown at $t=1500$ using linear reconstruction
        (red triangles), WENO (green circles) and PPM (blue squares).
        Refinement levels are marked by coloured boxes (top panel),
        with red corresponding to the base level grid.
        The analytical solution of Eq. \eqref{Eq:DiffEq} is
        plotted for comparison while a close-up view of the interface
        between the first two grid levels is shown in the bottom panel.
	}
	\label{fig:DiffEqAMR}
  \end{figure}

 The limiting scheme in Eq. \eqref{Eq:RadSpeedLim} depends on the
 optical depth of individual cells, which is inversely proportional
 to the resolution. Therefore, when AMR is used, there can be situations
 where this limiting is applied in the coarser levels, but not in the
 finer ones. Furthermore, when using HLLC, the solver
 is replaced by HLL for every zone where Eq. \eqref{Eq:RadSpeedLim}
 is enforced. To study the code's performance under these conditions,
 we have run this test on a static AMR grid using $128$ zones at the coarsest
 level with 6 levels of refinement with a jump ratio
 of 2, yielding an equivalent resolution of $8192$ zones.
 We choose $\sigma=50$ so that levels 0 to 4 are solved with the
 HLL solver limiting the maximum signal speeds accordingly to Eq.
 \eqref{Eq:RadSpeedLim}, while levels 5 and 6 are solved using the
 HLLC solver.
 The solution thus obtained converges to the analytic
 solution of Eq. \eqref{Eq:DiffEq} in all the refinement levels
 (see Fig. \ref{fig:DiffEqAMR}).

 However, we have observed the formation of spurious overshoots at the
 boundaries between refinement levels.
 These artifacts are drastically reduced if the order of the reconstruction
 scheme is increased; for instance, if the weighted essentially
 non-oscillatory (WENO) method by \cite{JiangShu} or the
 piecewise parabolic method (PPM)
 by \cite{PPM} are used, as shown in Fig. \ref{fig:DiffEqAMR}. 
 We argue that such features, which are not uncommon in AMR
 codes \citep{Choi_etal2004, Chilton_Colella2010}, can be attributed to the
 refluxing process needed to ensure correct conservation of momentum and
 total energy \citep[see][]{AMRPLUTO}.
 In addition, the presence of source terms requires additional care
 when solving the Riemann problem betwen fine-coarse grids due to temporal
 interpolation  \citep{Berger_LeVeque1998}.
 We do not account here for such modifications and defer these potential
 issues to future investigations.

%%%%%%%%%%%%%%%%%%%%%%%%%%%%%%%%%%%%%%%%%%%%%%%%%%%%%%%%%%%%%%%%%%%%%%%%
 \subsection{Shadows}\label{S:Shadows}
%%%%%%%%%%%%%%%%%%%%%%%%%%%%%%%%%%%%%%%%%%%%%%%%%%%%%%%%%%%%%%%%%%%%%%%%
 
 One of the main features of the M1 closure is its ability to
 reproduce the behavior of physical systems in which the angular
 distribution of the radiation specific intensity has strong
 spatial variations. One such example is
 a system where a free-streaming radiation field encounters a
 highly opaque region of space, casting a shadow behind it.
 To test the code performance when solving such
 problems, we have performed a test in
 which a shadow is formed behind a high-density elliptic cylinder,
 following \citet{HayesNorman2003} and using the same parameters
 as in \citet{Gonzalez2007}.
 
 Computations are carried out in the two-dimensional domain given by
 \mbox{$\left\lbrace(x,y)\in [-0.5,0.5]
 \,\mbox{cm}\times[0,0.6]\, \mbox{cm}\right\rbrace$}.
  Reflective boundary conditions are imposed at $y=0$. 
 A constant density \mbox{$\rho_0=1$ g cm$^{-3}$}
 is fixed in the whole
 space, except in the elliptic region,
 where $\rho=$ \mbox{$\rho_1=10^3$ g cm$^{-3}$}. In order to have a
 smooth transition between $\rho_0$ and $\rho_1$, the initial
 density field is defined as
 \begin{equation}
 \rho\,(x,y)=\rho_0 + \frac{\rho_1-\rho_0}{1+e^\Delta},
 \end{equation}
 where
 \begin{equation}
 \Delta = 10 \left[
 \left(\frac{x}{x_0}\right)^2 +\left(\frac{y}{y_0}\right)^2 -1
 \right],
 \end{equation}
 with \mbox{$(x_0,y_0)=(0.10,0.06)$ cm}. In such a way, the region with
 $\rho=\rho_1$ is approximately contained in an ellipse of
 semiaxes $(x_0,y_0)$. Initially, matter is set in thermal
 equilibrium with radiation at a temperature $T_0=290$ K,
 and fluxes and velocities are initialized to zero.
 The absorption opacity
 in the material is computed according to Kramers' law, i.e.,
 $\kappa=\kappa_0\left(\frac{\rho}{\rho_0}\right)
 \left(\frac{T}{T_0}\right)^{-3.5}$, with
 \mbox{$\kappa_0 = 0.1$ g$^{-1}$cm$^2$}, while scattering
 is neglected. Therefore, the cylinder's optical thickness
 along its largest diameter is approximately
 $\tau \approx 2\times 10^4$, which means that its
 width exceeds $\tau\gg 1$ times the photons' mean free path in
 that region.
 On the contrary, above $y>y_0$,
 the optical thickness is $\tau=0.1$, so that 
 the exterior of the cylinder is transparent to
 radiation while its interior is opaque.
 
 Radiation is injected from the left boundary at a temperature  
 \mbox{$\left(c\,E_r/4\,\sigma_{SB}\right)^{1/4}=1740$ K} 
 $>T_0$, with a flux 
    $\mathbf{F}_r=c\, E_r\,\hvec{e}_x$. Hence, the radiation
 field is in initially in the free-streaming limit, and should
 be transported at the speed of light in the transparent regions.
 
   \begin{figure*}[t]
	\centering
	\includegraphics[width=0.65\textwidth]{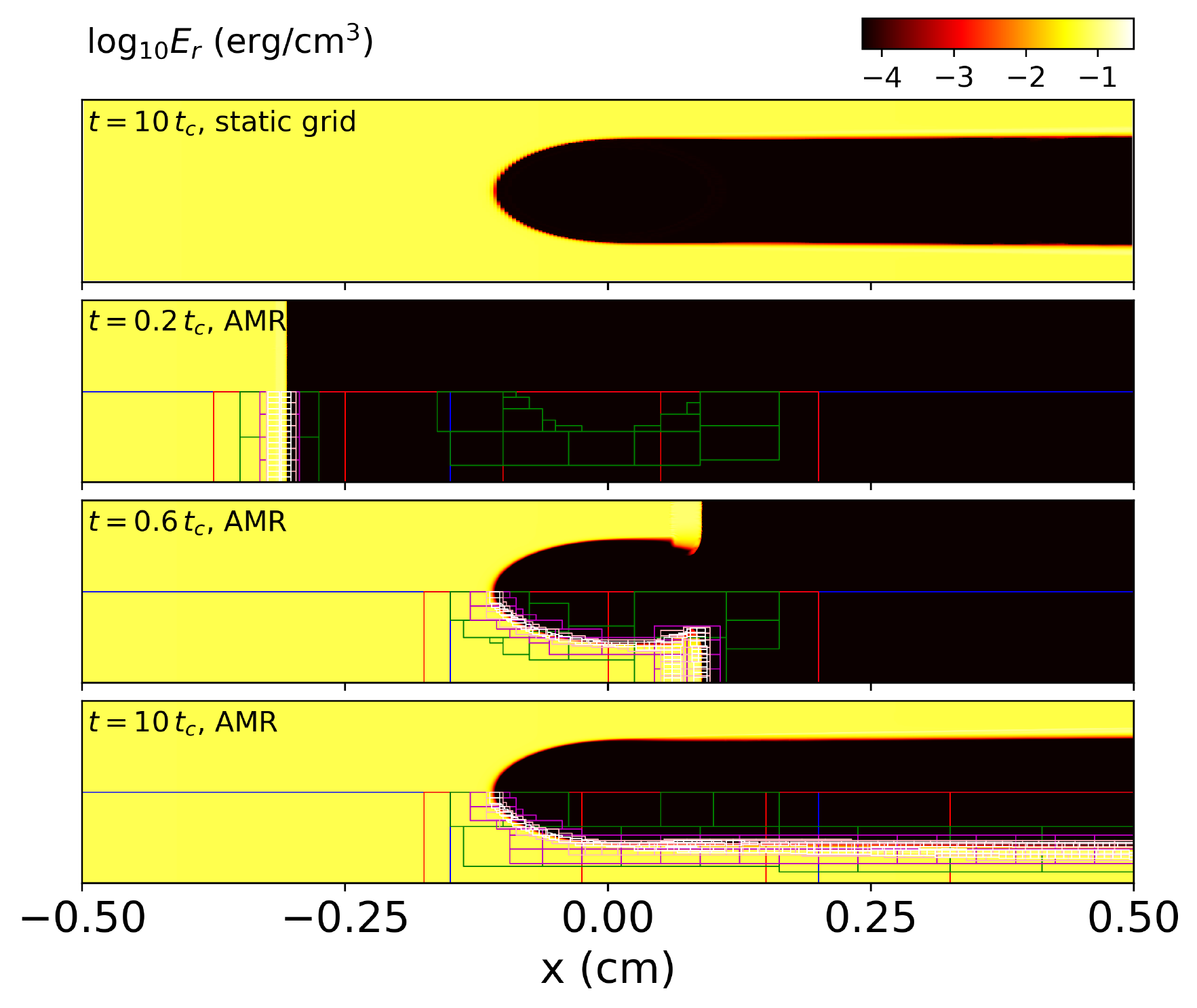}
	\caption{Radiation energy density maps obtained in the
	shadow test.
	The radiation front crosses the domain from left to right,
	casting a shadow behind an elliptic cylinder centered
	at $(0,0)$.
        From top to bottom we show the numerical solutions obtained, respectively,
	on a static uniform grid with resolution $280\times80$ at $t=10\,t_c$, 
        on the AMR grid ($80\times16$ zones on the base level) at
        $t=0.2\,t_c$, $0.6\,t_c$, and $10\,t_c$.
	The radiation front crosses the domain at the speed of light in the
        transparent regions.
	Refinement levels are superimposed with colored lines in the
	lower halves of these figures, corresponding to $l=0$ (blue), $1$ (red),
	$2$ (green), $3$ (purple), $4$ (golden), and $5$ (black),
	where $l$ is the refinement level.
        % It can be here seen that 
	%the highest $l$ values follow the sharpest variations in the
	%energy density.
  }
	\label{fig:ShadowTest}
  \end{figure*}  
 
 We have initially computed the system's evolution in a fixed
 uniform grid of resolution $280 \times 80$, using a fourth-order
 reconstruction scheme with a Courant factor $C_a=0.4$, and with
 $\Gamma=5/3$. 
 Simulations show a radiation front that crosses the space at light
 speed from left to right, producing a shadow behind the
 cylinder. After interacting with it, the shadow settles
 into a final stable state that is
 ideally maintained until the matter 
 distribution is modified due to its interaction with radiation.
 The radiation energy density distribution is shown in
 the upper panel of Fig. \ref{fig:ShadowTest} at
 $t=10\,t_c$, where
 \mbox{$t_c=1\,\mbox{cm}/c=3.336\times 10^{-11}$ s}
 is the light-crossing time, namely, the time it takes light
 to cross the domain horizontally in the transparent region.
 Behind the cylinder, radiation energy is roughly equal to its
 initial equilibrium value of
   $(4\,\sigma_{SB}/c)\,T_0^4$.  This value
 is slightly affected by small waves that are produced in the
 upper regions of the cylinder, where the matter distribution stops
 being opaque to radiation along horizontal lines.
 Above the cylinder, the radiation field remains
 equal to the injected one.
 The transition between the shadowed and transparent regions
 is abrupt, as it can be seen in Fig. \ref{fig:ShadowTest}.
 The shape of the $E_r$ profile along vertical cuts
 is roughly maintained as radiation is transported
 away from the central object.
 
 When IMEX-SSP2(2,2,2) is used, we have noticed that $E_r$ goes frequently
 below $0$ on the left edge of the cylinder where the radiation
 field impacts it.
 Still, the obtained solutions are stable
 and convergent as long as $E_r$ is floored to a small value
 whenever this occurs.
 As in Section \ref{S:PulseThick},
 the radiation flux is much smaller in those zones than both its
 flux and the source terms, and the problem does not occur if
 IMEX1 is used.
 
 We have used this same problem to test the code's
 performance when AMR is used  in a multidimensional setup.
 In this case, we have run the same simulation, using an initially
 coarse grid of resolution $80 \times 16$ set to adapt to changes
 in $E_r$ and $\rho$ \citep[see][]{AMRPLUTO}.
 We have used $5$ refinement levels, in every case with a
 grid jump ratio of $2$, which gives an equivalent
 resolution of $2560 \times 512$. The resulting energy profiles
 are plotted  in the lower panels  of Figure 
 \ref{fig:ShadowTest}, for $t=0.2\,t_c$, $0.6\,t_c$, and $10\,t_c$,
 and agree with those computed using a fixed grid.
   In each panel we have superimposed the refinement level. 

%%%%%%%%%%%%%%%%%%%%%%%%%%%%%%%%%%%%%%%%%%%%%%%%%%%%%%%%%%%%%%%%%%%%%%%%
 \subsection{Magnetized cylindrical blast wave}\label{S:TestRMHD}
%%%%%%%%%%%%%%%%%%%%%%%%%%%%%%%%%%%%%%%%%%%%%%%%%%%%%%%%%%%%%%%%%%%%%%%%
 
  \begin{figure*}[t]
	\centering
	\includegraphics[width=0.97\textwidth]{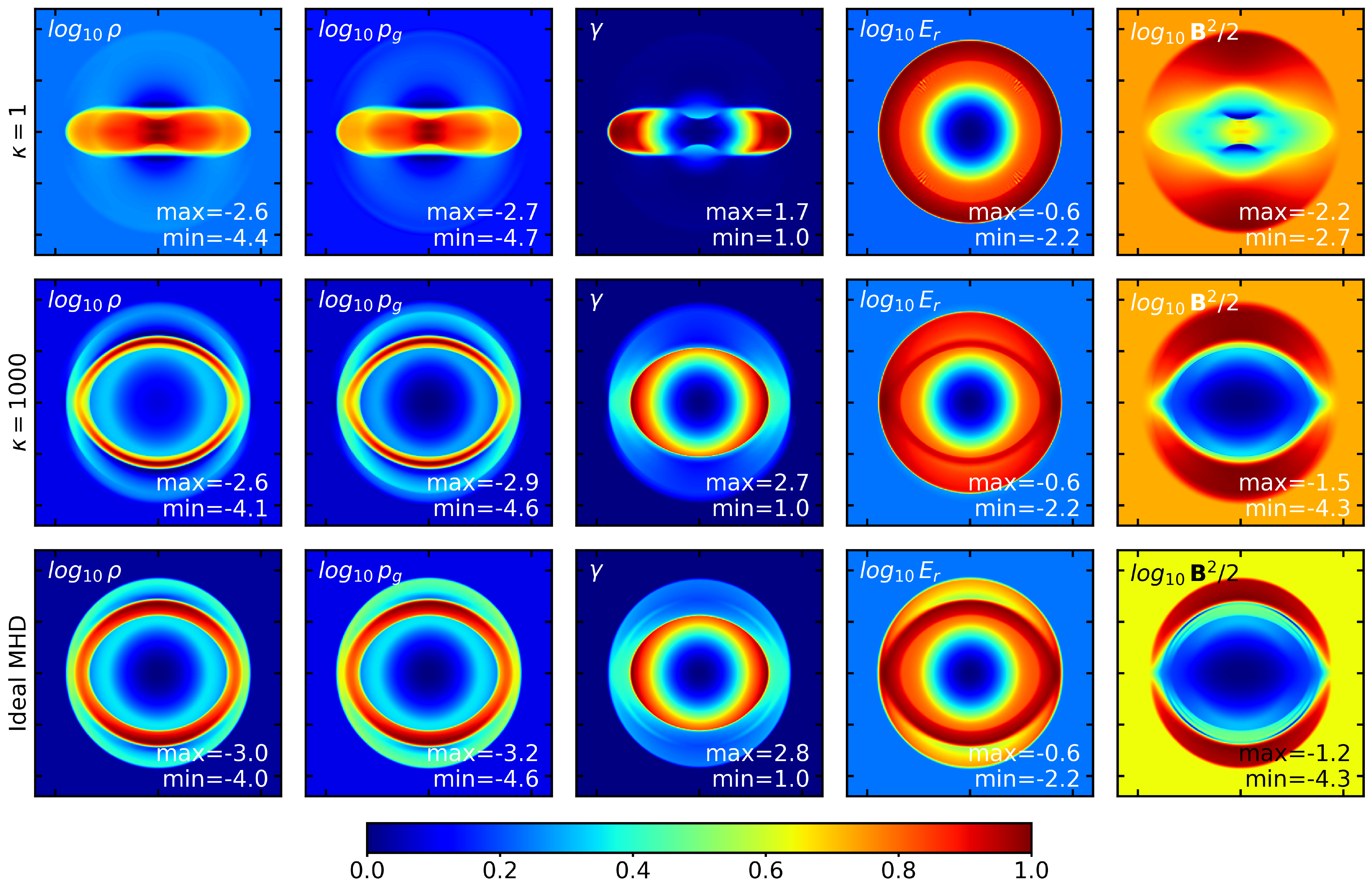}
	\caption{Density maps at $t=4$ in the magnetized cylindric
	blast wave test, corresponding to $\kappa=1$ (top row),
	$\kappa=1000$ (middle row) and ideal relativistic MHD (bottom row).}
	\label{fig:CylBlastWave}
  \end{figure*}
 
 We now examine a case in which matter is affected by both radiation
 and large-scale EM fields. We consider the case of a cylindrical
 blast wave, represented in a two-dimensional Cartesian grid as in
 the planar configurations described in Section \ref{S:PulseThin}. 
 In the context of MHD,
 this kind of problem has been used formerly to check the robustness
 of the employed methods
 when handling relativistic magnetized shocks, as well
 as their ability to deal with different kinds of
 degeneracies \citep[see e.g.][]{Komissarov,MignoneBodo2006}.
 In our case, we draw on this configuration as
 an example system that can switch from
 radiation-dominated to magnetically dominated regimes,
 depending on the material's opacity.
 To this end, we set up a cylindrical explosion
 from an area where the magnetic pressure is of the same order
 of the gas pressure, and both are
 smaller than the radiation pressure.
 Under this condition,
 matter dynamics is magnetically dominated
 when the opacities are low, and radiation-dominated in the
 opposite case. The latter case also serves to investigate the
 high-absorption regime in which both
 the diffusion approximation and LTE are valid.
 
 We consider a square domain defined as $(x,y)
 \in [-6,6]\times[-6,6]$,
 initially threaded by a uniform magnetic field,
 $\mathbf{B}=B_0\,\hvec{e}_x$ with $B_0=0.1$.
 Gas pressure and density are initially set as follows:
 \begin{equation}
    \left(\begin{array}{c}
      p  \\
     \rho   \end{array}\right) =
    \left(\begin{array}{c}
      p_1  \\
     \rho_1   \end{array}\right) \delta
     +
    \left(\begin{array}{c}
      p_0  \\
     \rho_0   \end{array}\right) (1-\delta)
 \end{equation}
 where $p_0 = 3.49\times 10^{-5}$, $\rho_0 = 10^{-4}$ are the ambient values
 while $p_1 = 1.31\times 10^{-2}$, $\rho_1 = 10^{-2}$ identify the
 over-pressurized region.
 Here $R=\sqrt{x^2 + y^2}$ is the cylindrical radius while $\delta\equiv\delta(R/R_0)$ is a
 taper function that decreases linearly for $R_0<R\le1$ (we use $R_0=0.8$).
 The ideal equation of state with $\Gamma = 4/3$ is used throughout the
 computations.

  A radiation field is introduced initially in equilibrium with the gas.
 Since $\mathbf{v}=\mathbf{0}$ in the whole domain, the condition of
 LTE is initially satisfied if $E_r=4\pi B(T)$ and
 $\mathbf{F}_r=\mathbf{0}$.
 These conditions are chosen in such a way
 that, close to the center of the domain,
 $p_g\sim\mathbf{B}^2/2<E_r/3$, where $\mathbf{B}^2/2$ gives the
 initial magnetic contribution to the total pressure
 (see Eq. \eqref{Eq:prstot}). To guarantee the condition
 $\nabla \cdot \mathbf{B}=0$, necessary for the solutions'
 stability, we have implemented in every case the constrained
 transport method.
 
 Figure \ref{fig:CylBlastWave} shows a set of 2D color maps
 representing the fields' evolution at $t=4$,
 using a resolution of $360\times 360$ zones.
 The two upper rows correspond to computations
 using $\sigma=0$ and $\kappa=1$ (top) or $1000$ (middle).
 For $\kappa=1$, the initial optical depth along the central sphere
 is $\tau\approx\rho_1 \kappa \Delta x= 0.02 \ll 1$,
 and therefore the material's expansion
 should not be noticeably affected by the radiation field.
 Indeed, in this case, the radiation energy
 profile expands spherically as in Section \ref{S:PulseThin}.
 The dynamic is magnetically dominated and matter is accelerated up
 to $\gamma\sim 1.7$ along the magnetic field lines along the $x$ axis,
 which is why the hydrodynamic variables are characterized by
 an elongated horizontal shape.
 
 The second row of Fig. \ref{fig:CylBlastWave} shows analog
 results obtained with
 $\kappa=1000$, where $\tau \approx 20 \gg 1$. In this case,
 the interaction of the radiation field with the gas during its
 expansion produces a much more isotropic acceleration. This
 acceleration is still maximal along the $x$ direction, due to
 the joint contributions of the magnetic field and the radiation
 pressure. This is why the Lorentz factor is larger in this
 case, reaching $\gamma\sim 2.7$.
 Gas density and pressure reach their maxima along
 an oblated ring, instead of the previous elongated distributions
 obtained with $\kappa=1$.
 As shown in the same figures, the
 magnetic field lines are pushed away from the center as matter
 is radially accelerated, producing a region of high magnetic
 energy density around the area where $\gamma$ is the highest,
 and a void of lower magnetic energy inside. Also differently from
 the previous case, the radiation energy distribution is no longer
 spherically symmetric due to its interaction with the matter 
 distribution.
 
 For high values of $\rho\kappa$, it is expected that the radiation
 reaches LTE with matter, as Eqs. \eqref{Eq:Gc}, 
 \eqref{Eq:RadRMHD1} and \eqref{Eq:RadRMHD2} lead to
 $\tilde{E}_r\rightarrow 4\pi B(T)$ and $\tilde{F}^i_r\rightarrow 0$
 for smooth field distributions that do not vary abruptly in time.
 In this limit, Eqs. \eqref{Eq:RadRMHD}-\eqref{Eq:RadRMHD2}
 can be reduced to those of relativistic MHD, redefining the total gas
 pressure as
 \begin{equation}
 p_{\rm tot}=p_g'+\frac{\mathbf{E}^2+\mathbf{B}^2}{2},
 \end{equation}
 with $p_g'=p_g+\tilde{p}_r$, and the enthalpy density as
 \begin{equation}
 \rho h_{\rm tot}=\rho h_g + \tilde{E}_r + \tilde{p}_r,
 \end{equation}
 where $\tilde{P}_r^{ij}=\tilde{p}_r\, \delta^{ij}$, which follows from
 the M1 closure in this limit. Taking a constant-$\Gamma$ EoS with
 $\Gamma=4/3$ in every case, the equations of state of both
 systems of equations coincide in the large-opacity limit, and
 therefore the results obtained with both of them are comparable.
 
\begin{figure}[t]
	\centering
	\includegraphics[width=0.47\textwidth]{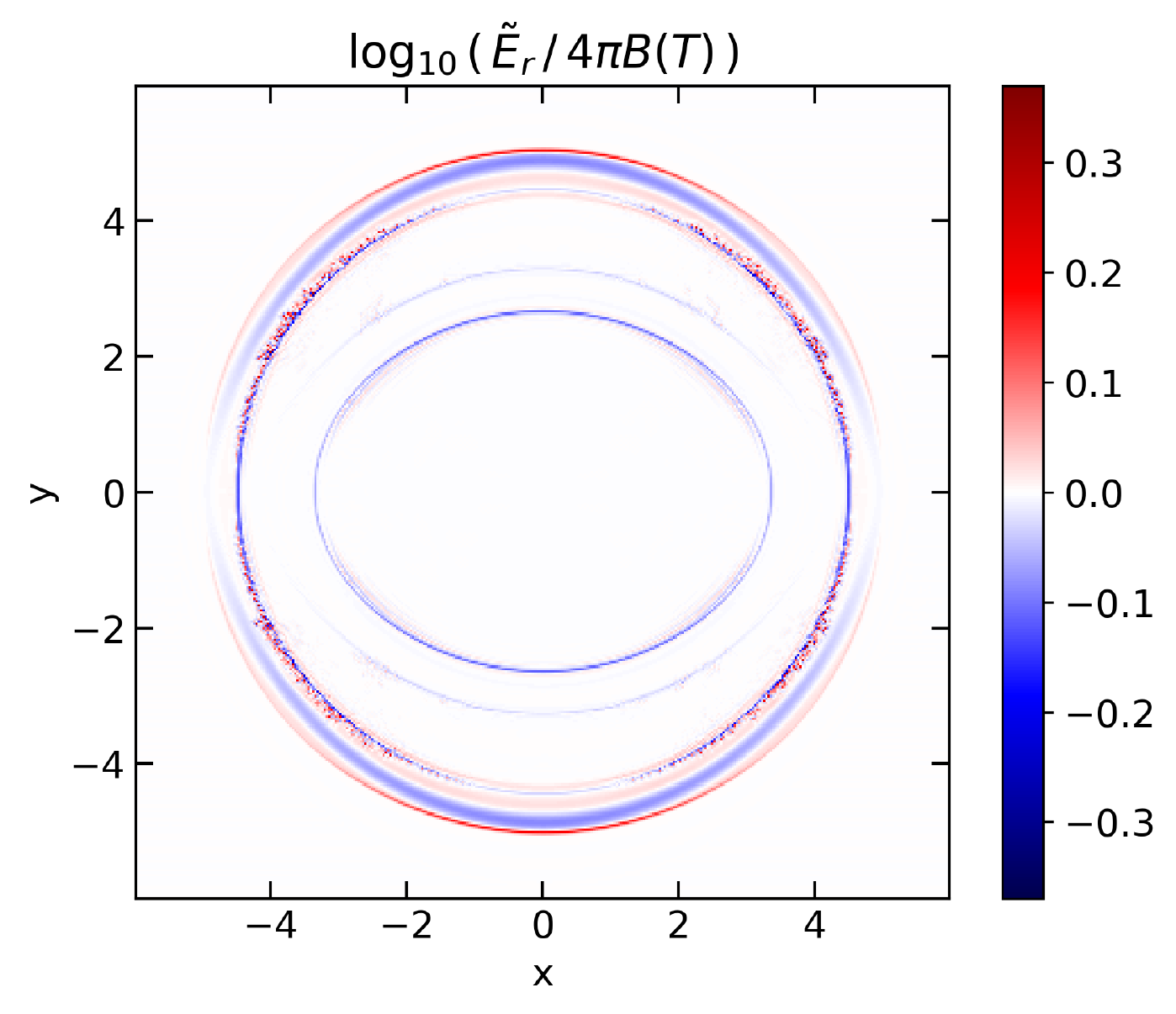}
	\caption{ Values of $\log_{10}\left(\tilde{E}_r/4\pi B(T)\right)$ 
	in the cylindric blast wave for $\kappa=1000$, computed at
	at $t=4$. The condition for LTE is here verified except in
	the closest regions to the shocks
	(see Fig. \ref{fig:CylBlastWave}).}
	\label{fig:CylBlastWaveErBT}
  \end{figure}

 The third row of Fig. \ref{fig:CylBlastWave} shows the results
 of an ideal relativistic MHD simulation performed in such a way,
 using the same initial conditions as before.
 To compute the gas pressure represented herein,
 it was assumed that $p_g'\simeq \tilde{p}_r=4\pi B(T)/3$, from where
 it is possible to extract $T$ and then $p_g$.
 Following the  same idea, an effective
 $E_{r}$ was computed boosting its comoving value, assumed to
 be equal to $4\pi B(T)$, and taking
 $\tilde{F}^i_r=0$. The resulting plots thus obtained are
 in fact similar to those computed with $\kappa=1000$, with
 slight differences that can be explained taking into account that
 $\kappa$ has a finite value, and that close to the shocks 
 the fields do not satisfy
 the conditions of being smooth and varying slowly with time.
 Consequently, the value of $\tilde{E}_r$ can be different than
 $4\pi B(T)$ in the regions of space that are close to 
 discontinuities, which means that the hypothesis of LTE,
 assumed by ideal MHD, is not satisfied in the whole domain.
 This is verified in Figure \ref{fig:CylBlastWaveErBT}, where
 it is shown that, for $\kappa=1000$,
 the ratio $\tilde{E}_r/4\pi B(T)$ differs from $1$ only in the
 regions that are close to shocks, shown
 in Fig. \ref{fig:CylBlastWave}.

%%%%%%%%%%%%%%%%%%%%%%%%%%%%%%%%%%%%%%%%%%%%%%%%%%%%%%%%%%%%%%%%%%%%%%%%
 \subsection{Parallel Performance}\label{S:Scaling}
%%%%%%%%%%%%%%%%%%%%%%%%%%%%%%%%%%%%%%%%%%%%%%%%%%%%%%%%%%%%%%%%%%%%%%%%
  
  Parallel scalability of our algorithm has been investigated in strong
  scaling through two- and three-dimensional computations.
  For simplicity, we have chosen the (unmagnetized) blast wave problem
  of Section \ref{S:TestRMHD} with $\kappa=10$ leaving the remaining
  parameters unchanged. 
  For reference, computations have been carried out with and
  without the radiation field on a fixed grid of $2304^2$ (in 2D)
  and $288^3$ (in 3D) zones, a constant time step and the
  solver given by Eq. \eqref{Eq:LFR}.
  The number of processors - Intel Xeon Phi7250 (KnightLandings) at 1.4 GHz - 
  has been varied from $N_{\rm CPU}=8$ to $N_{\rm CPU} = 1024$.

\begin{figure}[t]
	\centering
	\includegraphics[width=0.47\textwidth]{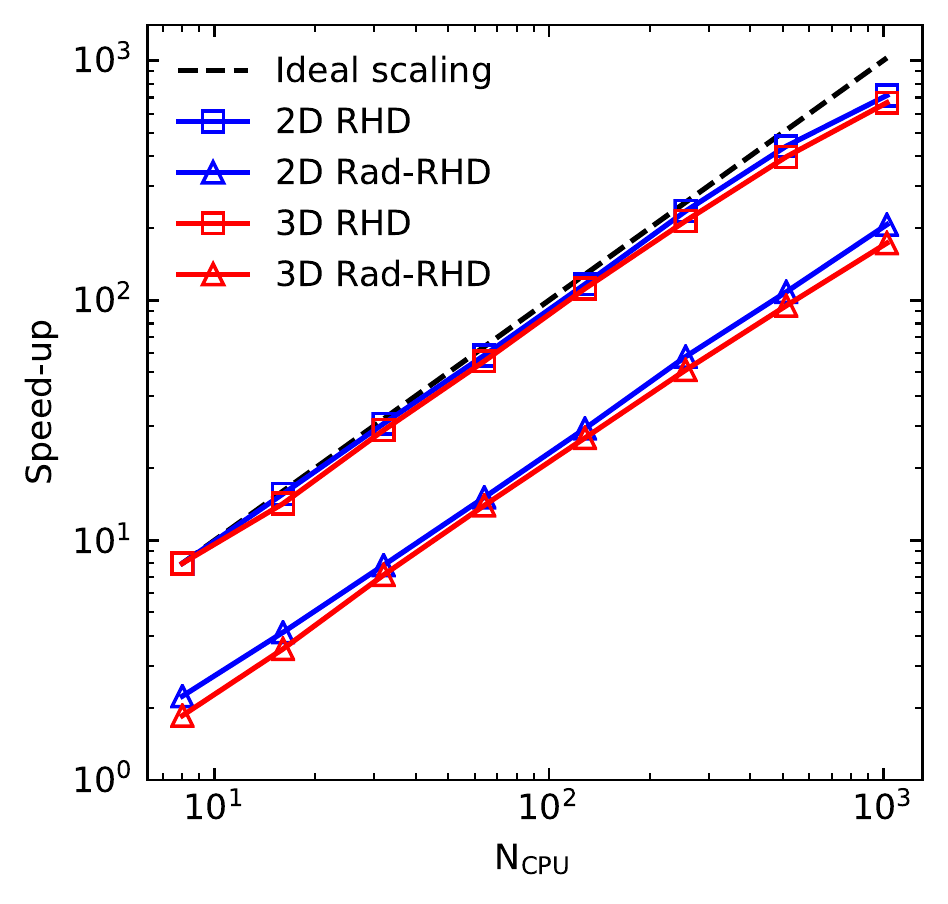}
	\caption{ Speed-up for the 2D (blue) and 3D (red) blast wave tests,
        with and without radiation fields (triangles and squares)
        as a function of the number of processors.
        The ideal scaling law (dashed, black line) is shown
	for comparison.}
	\label{fig:Scaling2D3D}
  \end{figure}

  The corresponding speed-up factors are plotted in Fig. \ref{fig:Scaling2D3D}
  as a function of N$_\mathrm{CPU}$ (solid lines with symbols) together with
  the ideal scaling-law curve $\propto N_{\rm CPU}$.
  We compute the speedup factors as $S = T_{\rm ref}/T_{\mathrm{N}_\mathrm{CPU}}$
  where $T_{\rm ref}$ is a normalization constant while $T_{\mathrm{N}_\mathrm{CPU}}$
  is the total running time for a simulation using $N_\mathrm{CPU}$
  processors.
  
  Overall, favorable scaling properties are observed in two and three
  dimensions with efficiencies that remain above $90\%$ up to 256 cores and
  drops to $\sim 70\%$ when $N_{\rm CPU} = 1024$.
  Slighlty better results are achieved when radiation is included, owing to
  the additional computational overhead introduced by the implicit part of the
  algorithm which uses exclusively local data without requiring additional
  communication between threads.

  Note that, for convenience, we have normalized the curves to the
  corresponding running time without the radiation field.
  This demonstrates that, by including radiation, the code is (approximately)
  four times more expensive than its purely hydro counterpart, regardless of the
  dimensionality.
    
 \section{Summary}\label{S:Summary}
 
 We have presented a relativistic
 radiation transfer code, designed to function
 within the \sftw{PLUTO} code.
 Our implementation can be used together with the relativistic HD
 and MHD modules of \sftw{PLUTO} to solve the equations of radiation
 transfer under the gray approximation.
 Integration is achieved through one of two possible IMEX schemes,
 in which source terms due to radiation-matter interaction
 are integrated implicitly and flux divergences, as well as every
 other source term, are integrated explicitly. 
 The transition between optically thick
 and thin regimes is controlled by imposing the M1 closure to the
 radiation fields, which allows to handle both the diffusion and
 free-streaming limits.
 Opacity coefficients can be arbitrarily defined, depending on
 problem at hand, as functions of the primitive variables.
 
 In our implementation, a novel HLLC Riemann solver for radiation
 transport has been introduced.
 The new solver is designed to improve the accuracy of the solutions
 with respect to it predecessors (such as HLL)
 in optically thin regions of space.
 The module has been designed to function with either Cartesian,
 cylindrical or spherical coordinates in multiple spatial dimensions
 and it is suitable for either serial or parallel computations. 
 Extension to adaptive grids, based on the standard implementation of
 the \sftw{CHOMBO} library within the code, has also been presented.
 
 We have performed a series of numerical benchmarks
 to assess the module performance under different configurations, 
 including handling of radiation transport,
 absorption, and emission in systems with different characteristics.
 Our results demonstrate excellent stability
 properties under the chosen parameters, in both the free-streaming
 and diffusion limits.
 In the latter case, numerical diffusion is
 successfully controlled by limiting the signal speeds of the
 radiation transport equations whenever the material is opaque
 across single grid cells.
 Overall, the transition between both regimes has been properly
 captured by the code in all the considered cases.
 For optically thin transport, our HLLC solver
 produces more accurate solutions when compared to HLL.
 Regarding the implemented IMEX schemes, we have seen a similar
 performance of both IMEX-SSP2(2,2,2) and IMEX1 except in tests where
 the order of magnitude of the radiation flux is much smaller
 than both its source terms and the divergence of its own flux,
 in which IMEX1 seems to have better stability and
 positivity-preserving properties. 
 When AMR is used, the obtained solutions 
 exhibit a similar overall behavior to those computed using a fixed
 grid. Good agreement is also shown with standard tests whenever the
 comparison is possible.  Furthermore, parallel performance tests
 show favorable scaling properties which are comparable to those of
 the RHD module of \sftw{PLUTO}.

 The code presented in this work will be made publicly available
 as part of future versions of \sftw{PLUTO}, which can currently
 be downloaded from \url{http://plutocode.ph.unito.it/}.
 \\\\

 \acknowledgments
 The authors would like to thank the referee for his/her constructive
 comments which helped to improve the quality of this manuscript.
 Also, we acknowledge the CINECA award under the ISCRA initiative, for
 the availability of high performance computing resources and support.

\appendix
 \section{Signal speeds}\label{S:AppSpeeds}

  We describe in this section the computation of the characteristic
  signal speeds used in the explicit step (see Section \ref{S:RSU}).
  In the particular form of Equations
  \eqref{Eq:RadRMHD}-\eqref{Eq:RadRMHD2}, the MHD fluxes are
  independent of the radiation variables, and vice-versa.
  Hence, the Jacobian matrices of the system are block-diagonal,
  one block corresponding to MHD and the other to radiation
  transport. Consequently, their corresponding
  sets of eigenvalues can be obtained
  by computing the eigenvalues of each of these blocks individually.
  
  For the MHD block, maximum and minimum signal speeds are computed
  as detailed in \cite{MignoneBodo2006} and \cite{MignoneMcKinney}.
  The treatment needed for the radiation wave speeds is rather
  simpler, as a short
  calculation shows that, for every direction
  $d$, the radiation block depends 
  only on the angle $\theta$ between $\mathbf{F}_r$ and
  $\hvec{e}_d$, and on
  $f=\vert\vert\mathbf{F}_r\vert\vert/E_r$.
  This simplifies the calculation, which can be
  performed analitically as shown in \cite{Audit2002} and
  \cite{Skinner2013}. The full set of eigenvalues of the
  radiation block, which we denote as
  $\{\lambda_{r1},\lambda_{r2},\lambda_{r3}\}$, can be computed
  as
  \begin{equation}
  \label{Eq:RadSigSpeed1}
  \lambda_{r1} = \frac{f\cos\theta -\zeta(f,\theta)}{\sqrt{4-3f^2}},
  \end{equation}
  \begin{equation}
  \label{Eq:RadSigSpeed2}
  \lambda_{r2} = \frac{3\xi(f)-1}{2f}\cos\theta,
  \end{equation}
  \begin{equation}
  \label{Eq:RadSigSpeed3}
  \lambda_{r3} = \frac{f\cos\theta +\zeta(f,\theta)}{\sqrt{4-3f^2}},
  \end{equation}
  where $\xi(f)$ is defined in Eq. \eqref{Eq:M13}, while
  \begin{equation}\label{Eq:RadSigSpeed4}
  \zeta(f,\theta) = 
  \left[
  	\frac{2}{3} \left(
  			4-3f^2-\sqrt{4-3f^2}
  		\right) 
  		  +  2\,\cos^2\theta \left(
			2-f^2-\sqrt{4-3f^2}  		
  		\right)
  	\right]^{1/2} .
  \end{equation}
  When $f=0$, $\lambda_{r2}$ is replaced by 0, i.e.,
  its limit when $f\rightarrow 0$.
  It can be seen from Equations
  \eqref{Eq:RadSigSpeed1}-\eqref{Eq:RadSigSpeed4} that the
  following inequalities hold for every value of
  $f$ and $\theta$:
  \begin{equation}
  \lambda_{1r}\leq\lambda_{r2}\leq\lambda_{r3}.
  \end{equation}
  In the free-streaming limit ($f=1$), all these eigenvalues coincide
  and are equal to $\cos\theta$,
  which gives $\lambda_{rj}= \pm 1$ in
  the parallel direction to $\mathbf{F}_r$, and $\lambda_{rj}=0$
  in the perpendicular ones for $j=1,2,3$.
  On the other hand, in the diffusion limit ($f=0$),
  we have \mbox{
  $(\lambda_{r1},\lambda_{r2},\lambda_{r3})
  =(-1/\sqrt{3},0,1/\sqrt{3})$}
  in every direction.
  
  The above analysis can be applied to homogeneous hyperbolic systems.
  Although the equations of Rad-RMHD do not belong to this
  category, this is not a problem when radiation transport
  dominates over radiation-matter interaction. On the contrary,
  in the diffusion limit, the moduli of the maximum and 
  minimum speeds, both equal to $1/\sqrt{3}$,
  may be too big and lead to an excessive numerical diffusion.
  In those cases, the interaction
  terms need to be taken into account to estimate the wave speeds.
  With this purpose, following \cite{Sadowski2013}, we include in the
  code the option of locally limiting the maximum and minimum speeds
  by means of the following transformations:
  \begin{equation}\label{Eq:RadSpeedLim}
  \begin{split}
    \lambda_{r,\,L}  & \rightarrow \max 
    \left( \lambda_{r,\,L} , -\frac{4}{3\tau} \right) \\
    \lambda_{r,\,R}  & \rightarrow \min 
    \left( \lambda_{r,\,R} , \frac{4}{3\tau} \right)
  \end{split},
  \end{equation}
  where $\tau=\rho\,\gamma\,\chi\,\Delta x$ is the optical
  depth along one cell, being $\Delta x$ its width in the
  current direction. Hence, this limiting is only applied whenever
  cells are optically thick. The reduced speeds in Eq.
  \eqref{Eq:RadSpeedLim} are based on a diffusion equation
  like Eq. \eqref{Eq:DiffEq},
  where the diffusion coefficient is $1/3\rho\chi$.

   \section{Semi-analytical proof of $\lambda_L\leq\lambda^*\leq\lambda_R$}
   \label{S:AppLambdaS}
   
   In order to check the validity of Equation \eqref{Eq:CondLambdaS},
   we have verified the following relations:
	\begin{align}\label{Eq:BA1}
	\lambda_R&\geq\max\left(\frac{B_R}{A_R},\frac{B_L}{A_L}\right)\\
	\label{Eq:BA2}
	\lambda_L&\leq\min\left(\frac{B_R}{A_R},\frac{B_L}{A_L}\right).
	\end{align}
	As in Section \ref{S:HLLC}, we
	omit the subindex $r$, as it is understood that only radiation
	fields are here considered.
	
	We begin by proving the positivity of $A_R$. From its definition
	in Equation \eqref{Eq:Adef}, we have:
	\begin{equation}\label{Eq:ARER}
	\frac{A_R}{E_R}=\lambda_R-f_{x,R}
	=\max(\lambda_{3,L},\lambda_{3,R})-f_{x,R}\,,
	\end{equation}
	where $\lambda_{3,L/R}=\lambda_3(f_{L/R},\theta_{L/R})$.
	Since $E>0$, we can conclude from Eq. \eqref{Eq:ARER} that
	$A_R\geq 0$ is automatically satisfied if
	\begin{equation}\label{Eq:ARER2}
	\lambda_3(f,\theta)\geq f\cos\theta\,\,\forall\, (f,\theta)\,.
	\end{equation}
	From Eq. \eqref{Eq:RadSigSpeed3}, this condition can be
	rewritten as
	\begin{equation}
	\zeta(f,\theta)\geq f\cos\theta\,\left(\Delta-1\right)\,,
	\end{equation}
    where $\Delta = \sqrt{4-3f^2}$. Taking squares at both sides and
    rearranging terms, this condition reads
    \begin{equation} \label{Eq:ARER3}
    X(f) + Y(f) \cos^2\theta \geq 0,
    \end{equation}
	where $X(f) = \frac{2}{3}(4-3f^2-\Delta)$ and
	$Y(f) = (1-f^2)(4-3f^2-2\Delta)$. Since only the second of these
	terms can be smaller than $0$, it is enough to prove that
	\eqref{Eq:ARER3} holds for $\cos^2\theta=1$, since
	that yields the
	minimum value that the left-hand side can take when $Y<0$.
	Hence, it is enough to prove
	\begin{equation}
	X(f) + Y(f) = \frac{1}{3}\Delta^2 (5-3f^2-2\Delta) \geq 0\,,
	\end{equation}
	which holds since the last term between parentheses is always
	greater than or equal to $0$. This finishes the proof of Eq.
	\eqref{Eq:ARER2}. Using the same equations, we can see that
	$\lambda_3(f,\theta)-f_x = 0$ is only satisfied if $f=1$.
	An analog treatment	can be used for $A_L$,
	from which we arrive to the following inequalities:
	\begin{align}
	A_R\geq 0, \,\,&\mbox{and}\,\,A_R>0\,\,\forall\, f\in[0,1)\\
	A_L\leq 0, \,\,&\mbox{and}\,\,A_L<0\,\,\forall\, f\in[0,1).
	\end{align}

	We now proceed to verify Equations \eqref{Eq:BA1} and
	\eqref{Eq:BA2}, firstly considering the case $f_L,f_R<1$,
	in which $A_{L/R}\neq 0$.
	Under this condition, the ratio $B_S/A_S$	depends only on
	$(f_L,f_R,\theta_L,\theta_R)$ as
	\begin{equation}
	\frac{B_S}{A_S}\equiv \alpha(\lambda_S,f_S,\theta_S) =
	\frac{(\lambda_S-\lambda_{2,S})
	f_S\cos\theta_S-(1-\xi(f_S))/2}{\lambda_S-f_S \cos\theta_S},
	\end{equation}
	with $S=L,R$.
	In order to verify Eq. 	\eqref{Eq:BA1}, we must prove
	$\lambda_R\geq B_R/A_R$ and $\lambda_R\geq B_L/A_L$.
	Since $\lambda_R=\max(\lambda_{3,L},\lambda_{3,R})$, we can
	write the first of these conditions considering the cases
	$\lambda_R=\lambda_{3,R}$ and $\lambda_R=\lambda_{3,L}$,
	as
	\begin{align}\label{Eq:condlambda1}
	\lambda_{3,R} &\geq \alpha (\lambda_{3,R},f_R,\theta_R)
	\,\,\,\forall\, (f_R,\theta_R)\\\label{Eq:condlambda2}
	\lambda_{3,L} &\geq \alpha (\lambda_{3,L},f_R,\theta_R)
	\,\,\, \forall\, (f_R,\theta_R):\lambda_{3,R}<\lambda_{3,L}
	\,\,\forall \,\lambda_{3,L} \in [-1,1] .
	\end{align}
	The first of these can be verified from the graph of
	$\lambda_3(f,\theta)-\alpha (\lambda_3(f,\theta),f,\theta)$,
	where it can be seen that this function is always positive
	and tends to $0$ for $f\rightarrow 1$.
	Similarly, we have checked the second one numerically by
	plotting 
	\mbox{$\lambda_{3,L}-\alpha (\lambda_{3,L},f_R,\theta_R)$}
	under the
	condition $\lambda_{3,R}<\lambda_{3,L}$, taking multiple values
	of $\lambda_{3,L}$ covering the range $[-1,1]$.
	The condition $\lambda_R\geq B_L/A_L$ can be proven in a similar
	fashion, by considering the cases $\lambda_L=\lambda_{1,L}$ and
	$\lambda_L=\lambda_{1,R}$. Since $\lambda_R\geq\lambda_{3,L}$,
	it is enough to prove the following conditions:
	\begin{align}\label{Eq:condlambda3}
	\lambda_{3,L} &\geq \alpha (\lambda_{1,L},f_L,\theta_L)
	\,\,\,\forall\, (f_L,\theta_L)\\\label{Eq:condlambda4}
	\lambda_{3,L} &\geq \alpha (\lambda_{1,R},f_L,\theta_L)
	\,\,\, \forall\, (f_L,\theta_L):\lambda_{1,R}<\lambda_{1,L}
	\,\,\forall \,\lambda_{1,R} \in [-1,1] ,
	\end{align}	
	which can be verified in the same manner, finishing the proof
	of Eq. \eqref{Eq:BA1} for the case $f_L,f_R<1$.
	The same procedure
	can be used to prove the validity of Eq. \eqref{Eq:BA2}.

	Unlike the RHD case, the maximum and minimum
	eigenvalues do not satisfy $\lambda_L<0$ and $\lambda_R>0$. 
	However, studying the parabolae
	defined at both sides of Eq. \eqref{Eq:PLPR}, it can be shown
	that $\lambda^*$ is always contained between $B_R/A_R$ and
	$B_L/A_L$, regardless of the order of these two values and of
	the signs of $\lambda_L$ and $\lambda_R$. Hence,
	\begin{equation}
	\lambda^*\in \left[\min\left(\frac{B_R}{A_R},\frac{B_L}{A_L}\right),
	\max\left(\frac{B_R}{A_R},\frac{B_L}{A_L}\right)\right].
	\end{equation}
	Together with relations \eqref{Eq:BA1} and \eqref{Eq:BA2},
	this proves Eq. \eqref{Eq:CondLambdaS} for $f_L,f_R<1$.
	These results are also valid
	in the cases $f_L=1$ and $f_R=1$ whenever the $A$ functions
	differ from 0. Let us now assume $f_a=1$ and $f_b\neq 1$.
	From Eqs. \eqref{Eq:Adef} and \eqref{Eq:Bdef},
	we have $A_a\cos\theta_a=B_a$ and consequently 
	$A_a=0$ implies that $B_a=0$. If $A_a=0$ and $A_b\neq0$,
	from \eqref{Eq:PLPR} we can extract that $\lambda^*=B_b/A_b$.
	Finally, from the above analysis, we know that 
	$\lambda_L\leq B_b/A_b\leq\lambda_R$, from which we conclude that
	\eqref{Eq:CondLambdaS} holds even in this case. The only
	remaining case is that in which $f_L=f_R=1$ and $A_L=A_R=0$,
	already considered in Section \ref{S:HLLC}, where the HLLC
	solver is replaced by the usual HLL solver.

\bibliography{refs}

\end{document}